\documentclass{aastex61}
\usepackage{multirow}
\usepackage{natbib}
\usepackage{array}
\usepackage{float}
\usepackage{xcolor}
\usepackage{graphicx}
\usepackage{epstopdf}
\usepackage[caption=false]{subfig}

\newcolumntype{L}[1]{>{\raggedright\let\newline\\\arraybackslash\hspace{0pt}}m{#1}}
\newcolumntype{C}[1]{>{\centering\let\newline\\\arraybackslash\hspace{0pt}}m{#1}}
\newcolumntype{R}[1]{>{\raggedleft\let\newline\\\arraybackslash\hspace{0pt}}m{#1}}
\definecolor{myred}{rgb}{1.0,0.40,0.40}

\newcommand{\kms}{km\,s$^{-1}$}

\received{\today}
\revised{xxxxxxxx xx, xxxx}
\accepted{xxxxxxxx xx, xxxx}
\submitjournal{ApJ}

\shorttitle{Coronal dimmings}
\shortauthors{Vanninathan et al.}

\begin{document}

\title{Plasma diagnostics of coronal dimming events}

\correspondingauthor{Astrid  M. Veronig}
\email{astrid.veronig@uni-graz.at}

\author{Kamalam Vanninathan}
\affil{Institute of Physics, University of Graz, 8010 Graz, Austria}

\author{Astrid M. Veronig}
\affil{Institute of Physics, University of Graz, 8010 Graz, Austria}
\affil{Kanzelh\"{o}he Observatory for Solar and Environmental Research,
University of Graz,
9521 Treffen, Austria}

\author{Karin Dissauer}
\affil{Institute of Physics, 
University of Graz, 
8010 Graz, Austria}

\author{Manuela Temmer}
\affil{Institute of Physics, University of Graz, 8010 Graz, Austria}



\begin{abstract}
Coronal mass ejections (CMEs) are often associated with coronal dimmings, i.e. transient dark regions that are most distinctly observed in Extreme Ultra-violet (EUV) wavelengths. Using Atmospheric Imaging Assembly (AIA) data, we apply Differential Emission Measure (DEM) diagnostics to study the plasma characteristics of six coronal dimming events. In the core dimming region, we find a steep and impulsive decrease of density with values up to 50--70\%. Five of the events also reveal an associated drop in temperature of 5--25\%. 
The secondary dimming regions also show a distinct decrease in density, but less strong, decreasing by 10--45\%. In both the core and the secondary dimming the density changes are much larger than the temperature changes, confirming that the dimming regions are mainly caused by plasma evacuation.
In the core dimming, the plasma density reduces rapidly within the first 20--30\,min after the flare start, and does not recover for at least 10\,hrs later, whereas the secondary dimming tends to be more gradual and starts to replenish after 1--2\,hrs. The pre-event temperatures are higher in the core dimming (1.7--2.6 MK) than in the secondary dimming regions (1.6--2.0 MK). Both core and secondary dimmings are best observed in the AIA 211\,\AA\ and 193\,\AA\ filters. These findings suggest that the core dimming corresponds to the footpoints of the erupting flux rope rooted in the AR, while the secondary dimming represents plasma from  overlying coronal structures that expand during the CME eruption.
\end{abstract}




\section{Introduction}
\label{sec:intro}

Coronal Mass Ejections (CMEs) are large-scale eruptions on the Sun often associated with flares \citep{Harrison1995, Zhou2003, Compagnino2017}. Substantial amounts of energy, plasma and magnetic field are released into the interplanetary medium during such ejections
\citep{Vourlidas2000, Webb2012}. Being the main drivers of severe space weather disturbances \citep{Schwenn2005, Lugaz2016, Riley2018}, CMEs are 
of high relevance for solar as well as space weather research and predictions \cite[e.g.][]{Schwenn2006, Gopalswamy2009, Howard2014, Green2018}.

CMEs originating from regions located centrally on the solar disk are most likely to be directed towards Earth \citep{StCyr2000}. However, they are difficult to observe in the Thomson-scattered white light by coronagraphs on-board spacecraft in the Sun-Earth line, since their propagation direction is far away from the plane of sky, and the measurements may be strongly affected by projection effects \citep[e.g.][]{Gopalswamy2000}. For these reasons, there is also a need for indirect means of studying the early evolution of Earth-directed CMEs. 
\cite{Hansen1974} were the first to report intensity `dimmings' in coronameter images from Mauna
Loa occurring simultaneously with CMEs. They have an appearance similar to coronal holes and are thus also referred to as `transient coronal holes' 
\citep{Rust1983,Hewish1985}.
A great advance in observing coronal dimmings associated with CMEs has been made with the advent of the Yohkoh/Soft X-Ray Telescope and Solar and Heliospheric Observatory/Extreme Ultraviolet Imaging Telescope (SoHO/EIT)  
\citep{Hudson1996, Sterling1997, Zarro1999, Thompson2000}. Since then, many case studies and statistical studies were performed using imaging at EUV and SXR wavelengths, in order to better understand the physics of coronal dimmings and how they relate to the CME evolution \citep[e.g. ][]{Zhukov2004, Reinard2008,  Bewsher2008}.

Coronal dimmings occur in association with the early eruption phase of CMEs.  It is believed that such dimmings are a consequence of density decrease in the low corona due to mass carried out by the CME \citep{Hudson1998, Mason2014}. Coronal dimming regions can be used as indirect means to estimate the source location, width and mass of CMEs \citep{Harrison2003, Aschwanden2016}. While observations of  CMEs originating from regions close to the center of the solar disk are subject to substantial projection effects, the coronal dimming region is most clearly discernible here making them a good proxy for studying Earth-directed CMEs.

Coronal dimmings can be differentiated into two types: core and secondary dimmings. The core dimmings are localized regions that occur in pairs on opposite sides of the source active region, and are rooted  in opposite magnetic polarities. They are interpreted as a signature of the footpoints of the erupting flux rope \citep{Sterling1997, Webb2000}. In contrast, secondary dimmings appear more shallow, diffuse and widespread. They are assumed to be caused by the expansion of the overall CME structure formed, and thus to correspond to the spatial extent of the CME observed in coronagraph data \citep{Mandrini2007}.

Several spectroscopic results support the presence of outflowing plasma in the dimming regions \citep{Harra2001, Harra2007, Miklenic2011, Tian2012} with velocities decreasing with time \citep{Jin2009} and prevalent thermal line broadening \citep{Mcintosh2009, Chen2010}. Density decrease of 35--40\% within the dimming regions \citep{Cheng2012} and plasma flows to replenish it \citep{Landi2012} have been identified using Differential Emission Measure (DEM) analysis.

In this paper, we study the plasma characteristics of coronal dimmings using DEM analysis. For this purpose, we selected six dimming events associated with CME/flare events of different speeds and flare classes for detailed case studies.  The plasma evolution in the core and the secondary dimming regions are studied in high cadence over a period of 12\,hrs. The paper is organized as follows, in Section\,\ref{sec:obs} we describe our data and methods of analysis. The results are presented in Section\,\ref{sec:results} and discussed in Section\,\ref{sec:sum_disc}. Finally, in Section\,\ref{sec:concl} we give the summary and conclusions of this work. 

\section{Observations and Data analysis}
\label{sec:obs}

The six EUV filters of the Atmospheric Imaging Assembly \citep[AIA,][]{Lemen2012} instrument on-board the Solar Dynamics Observatory \citep[SDO,][]{Pesnell2012} are sensitive to plasma over a wide temperature range from $\approx$10$^{5}$ to above 10$^{7}$\,K which facilitates the use of their data for the purpose of DEM reconstruction. Many DEM codes have been developed in the past years to use with AIA data \citep{Aschwanden2013, Hannah2012, Hannah2013, Plowman2013, Cheung2015}. For this work the code developed by \cite{Hannah2013} has been used, henceforth referred to as HK code.
The HK code uses regularized inversion to reconstruct a DEM from broad/narrow band filter observations using the equation 
\begin{equation}
I_{\mathrm{filter}} = \int R_{\mathrm{filter}}(T)\,\phi(T)\,\mathrm{d}T,
\label{eq:DEM}
\end{equation}
where $I_{\mathrm{filter}}$ is the intensity measured in a particular filter, $R_{\mathrm{filter}}(T)$  is the temperature response function of the instrument, and $\phi(T)$ is the required DEM.

For our analysis, we first binned the AIA images to a size of 512 pixels $\,\times\,$ 512 pixels (from the original images of 4096$\,\times\,$4096)  while conserving flux and then cut to a sub-map around the AR where the dimming could be clearly seen. The binning increases the signal to noise ratio, which is especially relevant in the AIA channels sensitive to hotter plasma  where the counts are relatively low (94 and 131\,\AA), thus improving the DEMs derived.
From the DEM maps, we then calculated for each rebinned pixel the emission measure (EM), the DEM 
weighted average temperature ($\bar{T}$) and plasma density ($\bar{n}$) using the following expressions:
\begin{equation}
EM=\int \phi(T)\,\mathrm{d}T,
\label{eq:em}
\end{equation}
\begin{equation}
\bar{T}=\frac{\int \phi(T)\,T\,\mathrm{d}T}{\int \phi(T)\,\mathrm{d}T},
\label{eq:temp}
\end{equation}
\begin{equation}
\bar{n}=\sqrt{\frac{\int \phi(T)\,\mathrm{d}T}{h}},
\label{eq:density}
\end{equation}
where $h$ is the distance along the line-of-sight (LOS). For this study we have used $h$=60 Mm corresponding to the coronal scale height for a temperature of about 1\,MK \citep[see also ][]{Vanninathan2015}.
Note that each of our rebinned pixels corresponds to $8 \times 8$ original AIA pixels, i.e. covers a field of about $4\arcsec \times 4\arcsec$ on the solar disk.

\begin{table*}[htbp]
\begin{center}
\begin{tabular}{| c | c | c | c | c |}
\hline
Date & GOES flare class & Flare location & CME speed$^\ddagger$ & EUV wave$^\dagger$ \\
 & & & (\kms) & speed (\kms) \\
\hline
01 August 2010 & C3.2 & N13 E21 & 850 & 312 \\
\hline
21 June 2011 & C7.7 & N17 W21 & 719 & -  \\
\hline
06 September 2011& X2.1 & N14 W18 & 575 & 1246  \\
\hline
19 January 2012 & M3.2 & N28 E13 & 1120 & -  \\
\hline
09 March 2012 & M6.3 & N17 W13 & 950 & 689  \\
\hline
14 March 2012 & M2.8 & N14 E01 & 411 & 485  \\
\hline
\end{tabular}
\end{center}
\caption{List of coronal dimming events under study. $^\ddagger$CME speed from SOHO/LASCO catalog. $^\dagger$EUV wave speed  from \cite{Nitta2013}. 
\label{table:events}}
\end{table*}

For the current work we have selected six coronal dimming events, which were observed against the solar disk, for detailed case study. We made sure that the list represented events associated with flares of different classes, ranging from C3 to X2 (and located within $30^\circ$ from the central meridian), and CMEs of different speeds, ranging from $\approx$400--1100\,\kms. 
Four of the events are associated with an EUV wave \citep{Nitta2013}, indicating the presence of large amplitude magnetosonic waves initiated by the impulsive lateral CME expansion \citep[e.g.][]{Kienreich2009, Patsourakos2009a, Veronig2010, Long2017}.
The events selected along with information on the associated flare, CME and EUV wave are summarized in Table\,\ref{table:events}. The CME speed given is the linear speed derived in the LASCO field-of-view, as listed in the LASCO/SOHO CME catalogue (\url{https://cdaw.gsfc.nasa.gov/CME_list/}), the EUV wave speeds are from the EUV wave catalogue compiled in \cite{Nitta2013}.

\section{Results}
\label{sec:results}
Using the HK code we have constructed DEM maps for all the selected events. EM,  density and temperature maps were derived from the DEMs as given in Equations\,\ref{eq:em}--\ref{eq:density}. In addition, we derived base-ratio maps for each of the parameters using a pre-event image (30\,min before the flare) as the base image. These maps were then used to characterize the plasma properties in the dimming region and their changes during the event. The differences between the core and secondary dimmings were studied and compared for all the events. Below we demonstrate our analysis in detail on two sample events: the 06 September 2011 event which was associated with the strongest flare in our list and a CME and EUV wave with high speed, and the event on 14 March 2012 which was associated with a slow CME. Thereafter, we summarize the main results obtained for the other four events, and add the related figures in the Appendix.

\subsection{Event on 06 September 2011}
The coronal dimming event of 06 September 2011 was associated with a X2.1 class flare in NOAA Active Region 11283, peaking at 22:12\,UT, and a CME with a linear speed of 575\,\kms\ in LASCO data.\footnote{We note that in this event the linear speed of the CME as observed by STEREO-A (EUVI-COR1-COR2) on the limb ($v=990$\,\kms ) is much larger than the LASCO plane-of-sky speed
($v=575$\,\kms ), indicative of strong projection effects \citep{Dissauer2016}. }
There was also a strong EUV wave with a speed of 1250\,\kms\ associated with this event, studied in detail in \cite{Dissauer2016}.

\begin{figure}[bth]
\includegraphics[width=0.9\columnwidth]{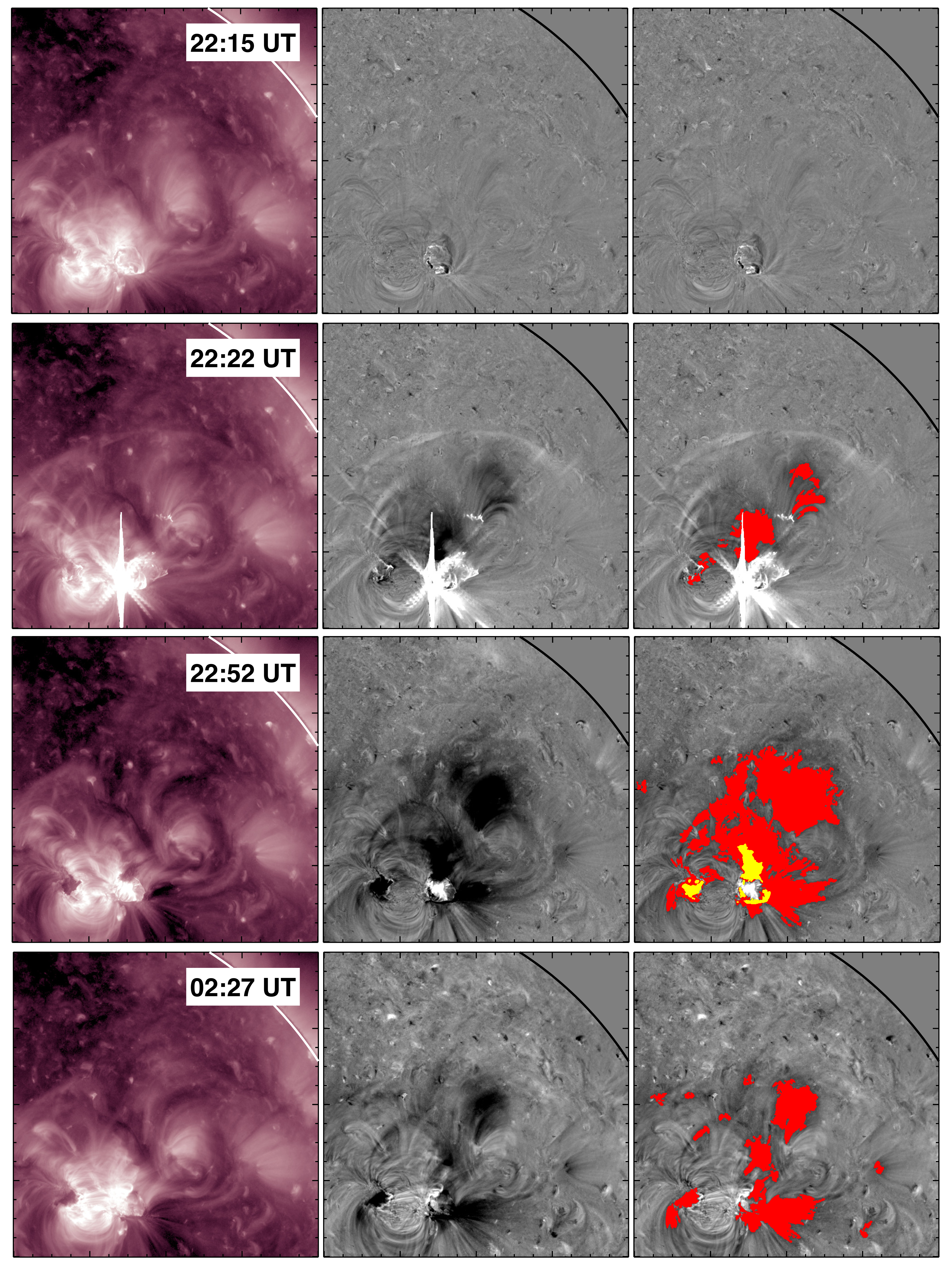}
\caption{Time evolution of the coronal dimming in SDO/AIA 211\,\AA~direct images (left) and logarithmic base-ratio images scaled from $-0.4$ to $+0.4$ (middle) for the event on 06 September 2011. In the right panels, we show the base ratio images overplotted with the instantaneously detected dimming regions (red pixels). The total extent of the core dimming region, as derived from the overall time series, is shown in the third panel (yellow pixels).}
\label{fig:dim_detection}
\end{figure}

\begin{figure}[htb]
\includegraphics[width=\columnwidth]{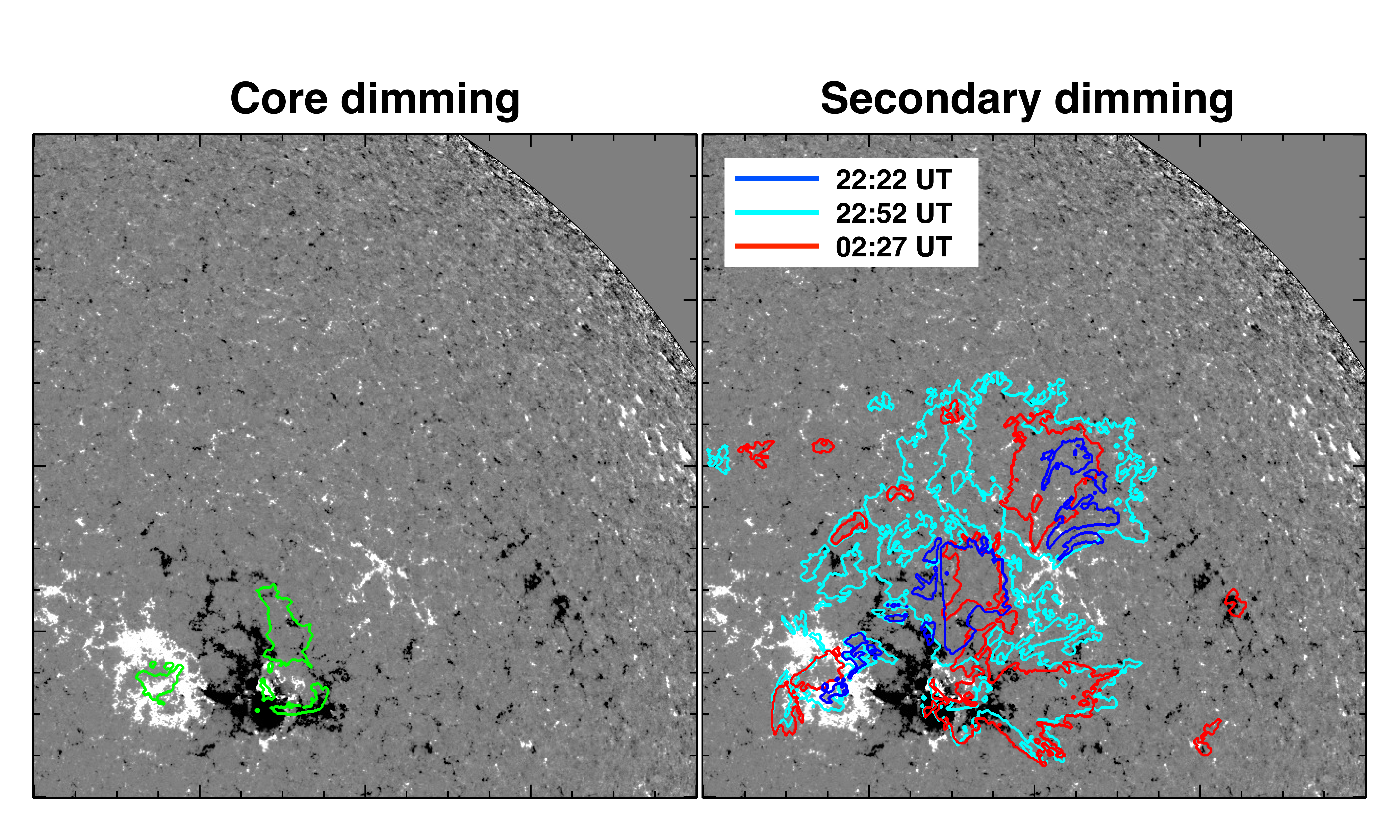}
\caption{SDO/HMI LOS magnetic field data (scaled to $\pm$100\,G) for the event on 06 September 2011 at a timestep close to the start of the flare.  Contours of the core (left) and secondary (right) dimming regions identified at different timesteps (color coded, see legend) are superimposed. The timesteps are the same as in Figure\,\ref{fig:dim_detection}.}
\label{fig:magnetic_field}
\end{figure}

To identify the core and secondary dimming regions, we use the aglorithm described in \cite{Dissauer2018}. The dimming detection is based on thresholding and region growing techniques applied to sequences of logarithmic base-ratio images. To identify locations of potential core dimmings within the overall dimming  detections,  
the pixels with the smallest intensities reached over the whole event duration are extracted from so-called minimum intensity maps. 
The results from the dimming detection are shown in Figure\,\ref{fig:dim_detection}. Here, the time evolution of the dimming region in the AIA 211\,\AA\ channel (left panels),  the corresponding logarithmically scaled base-ratio images (middle panels) as well as the dimming regions detected (right panels) are shown. Red (yellow) colors indicate the secondary (core) dimming regions identified. The first row in the image represents the state of the corona before the start of the event. Hence there is no dimming present yet. The second row shows the time step soon after the peak in the associated flare emission, where the dimming has already started to form. The third row shows an evolved dimming, with the number of dimming pixels increasing. This is close to the time step when the dimming area was at its maximum. In the last row, which is about 4\,hrs after the start of the associated flare, we see that the number of pixels detected as dimming has reduced and the area of the dimming region has already shrunk, indicating that the corona has started to be replenished.

We also compared the detected dimming region with photospheric magnetic field maps as shown in Figure\,\ref{fig:magnetic_field}. These are LOS magnetic field data from the Helioseismic and Magnetic Imager \citep[HMI;][]{Scherrer2012} onboard SDO. The core dimming regions (left panel) are localized, and do not show much expansion through the progress of the event. They occur in pairs of opposite polarity within the active region where the magnetic field density is high. This scenario conforms to the interpretation that these regions mark the footpoints of the erupting flux rope \citep{Sterling1997, Mandrini2005}. The secondary dimming region (right panel), on the other hand, first expands, reaches a maximum and thereafter shrinks with time, and is located predominantly in regions of weak magnetic fields.

\begin{figure*}[ht]
\includegraphics[width=\textwidth]{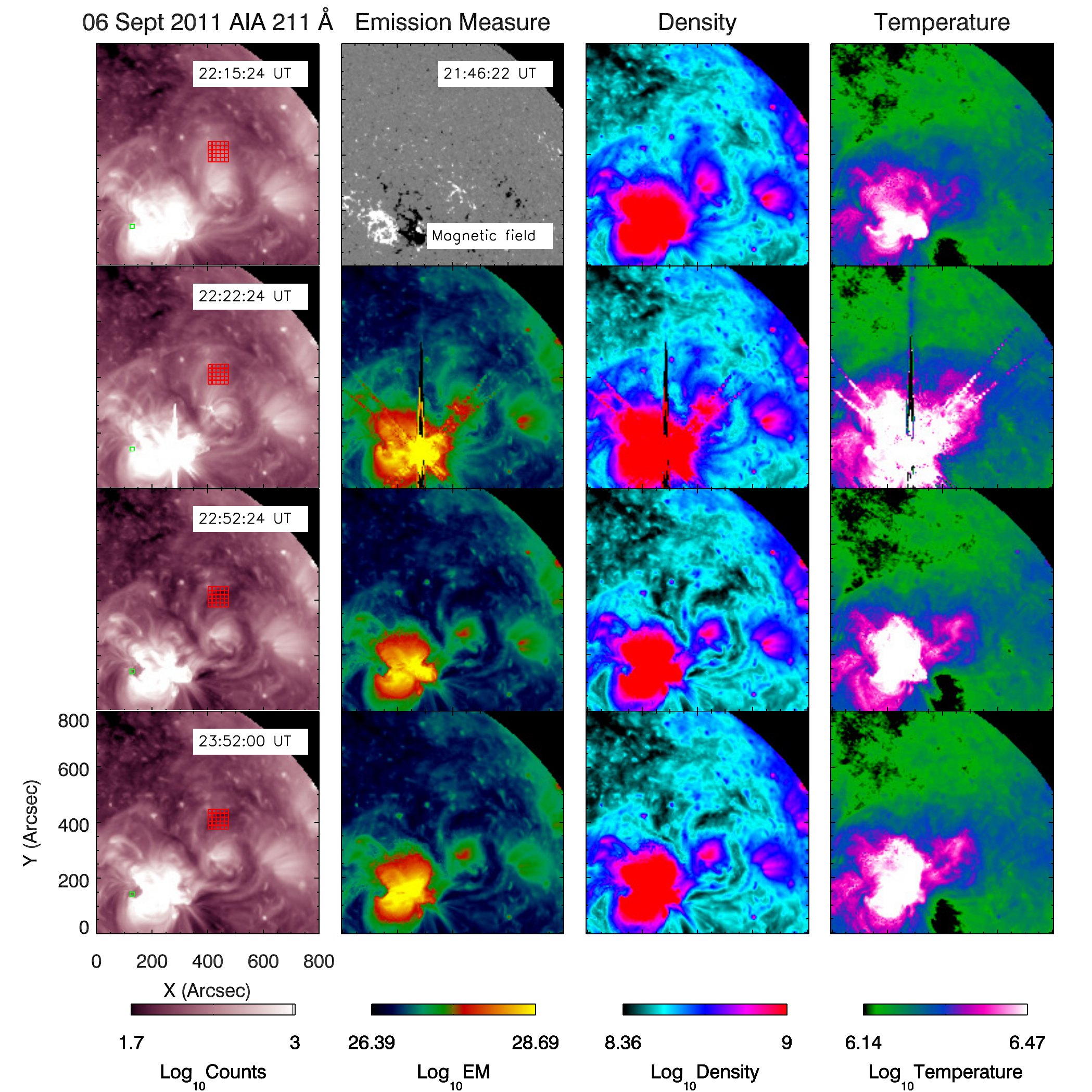}
\caption{Time evolution of SDO/AIA 211\,\AA\ images (first column), EM (second column), density (third column) and temperature (last column) for the event on 06 September 2011. The first panel of the second column shows the LOS magnetic field at a time close to the start of the flare scaled to $\pm$100\,G.}
\label{fig:plasma_evol_orig}
\end{figure*}

Using the HK code we constructed DEM maps from which we were able to derive EM, density and temperature maps. These maps are shown in Figure\,\ref{fig:plasma_evol_orig} together with the corresponding images from the AIA 211\,\AA\ filter. The first panel under EM is replaced with magnetic field data which can give us some information about the magnetic structure within the dimming region. We can see that as the dimming evolves, there is a significant decrease in EM and density. There is also a decrease observed in the temperature maps but this is not as substantial as the other quantities.

\begin{figure*}[ht]
\includegraphics[width=\textwidth]{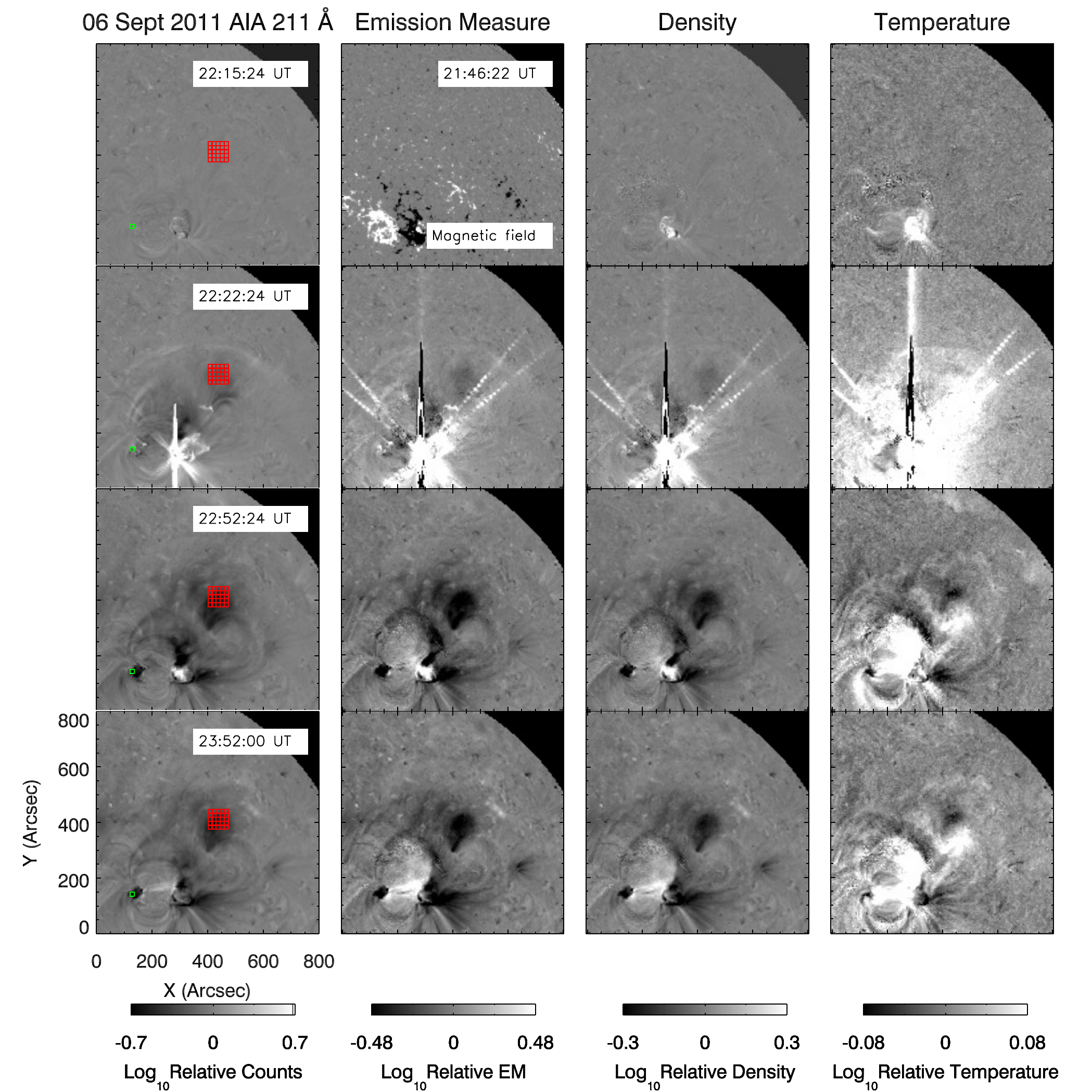}
\caption{Sequence of base-ratio maps visualizing the relative changes in the SDO/AIA 211\,\AA\ emission (first column), EM (second column), density (third column) and temperature (last column)  for the event on 06 September 2011. The base-ratio maps plotted here correspond to the direct maps shown in Figure\,\ref{fig:plasma_evol_orig}. The green and red boxes indicate a location within the core and secondary dimming region,  selected for detailed study.}
\label{fig:plasma_evol_baseratio}
\end{figure*}

To better see the changes in the derived quantities, we use base-ratio maps for enhancement. These images are shown in Figure\,\ref{fig:plasma_evol_baseratio}. Each panel in Figure\,\ref{fig:plasma_evol_baseratio} corresponds to the same panel in Figure\,\ref{fig:plasma_evol_orig}. From the dimming evolution in Figure\,\ref{fig:plasma_evol_baseratio} we clearly see a decrease in EM and density within the dimming region. The changes in these quantities closely reflect the changes seen in the AIA 211\,\AA\ channel. On the other hand, the temperature maps show a decrease only in certain small regions. In many pixels, especially those around the region of the strong X2.1 flare, we see an increase in temperature. This effect in the temperature reconstruction from the DEMs most probably arises from strong scattered light in these pixels, caused by the bright flaring regions.

\begin{figure*}
\centering
\begin{minipage}[t]{0.4\textwidth}
\includegraphics[width=6cm]{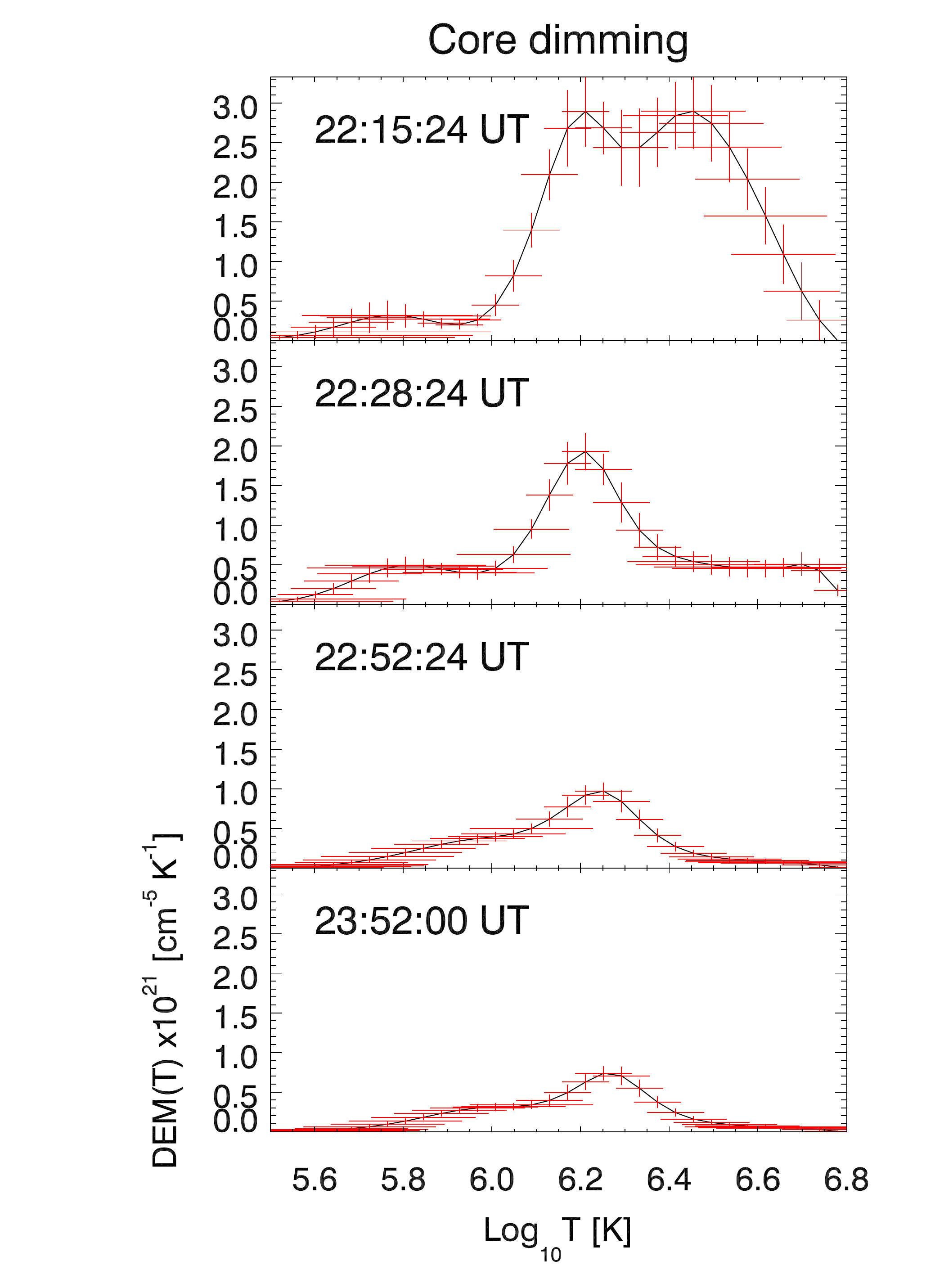}
\end{minipage}\hspace{-1cm}\begin{minipage}[t]{0.4\textwidth}~
\includegraphics[width=6cm]{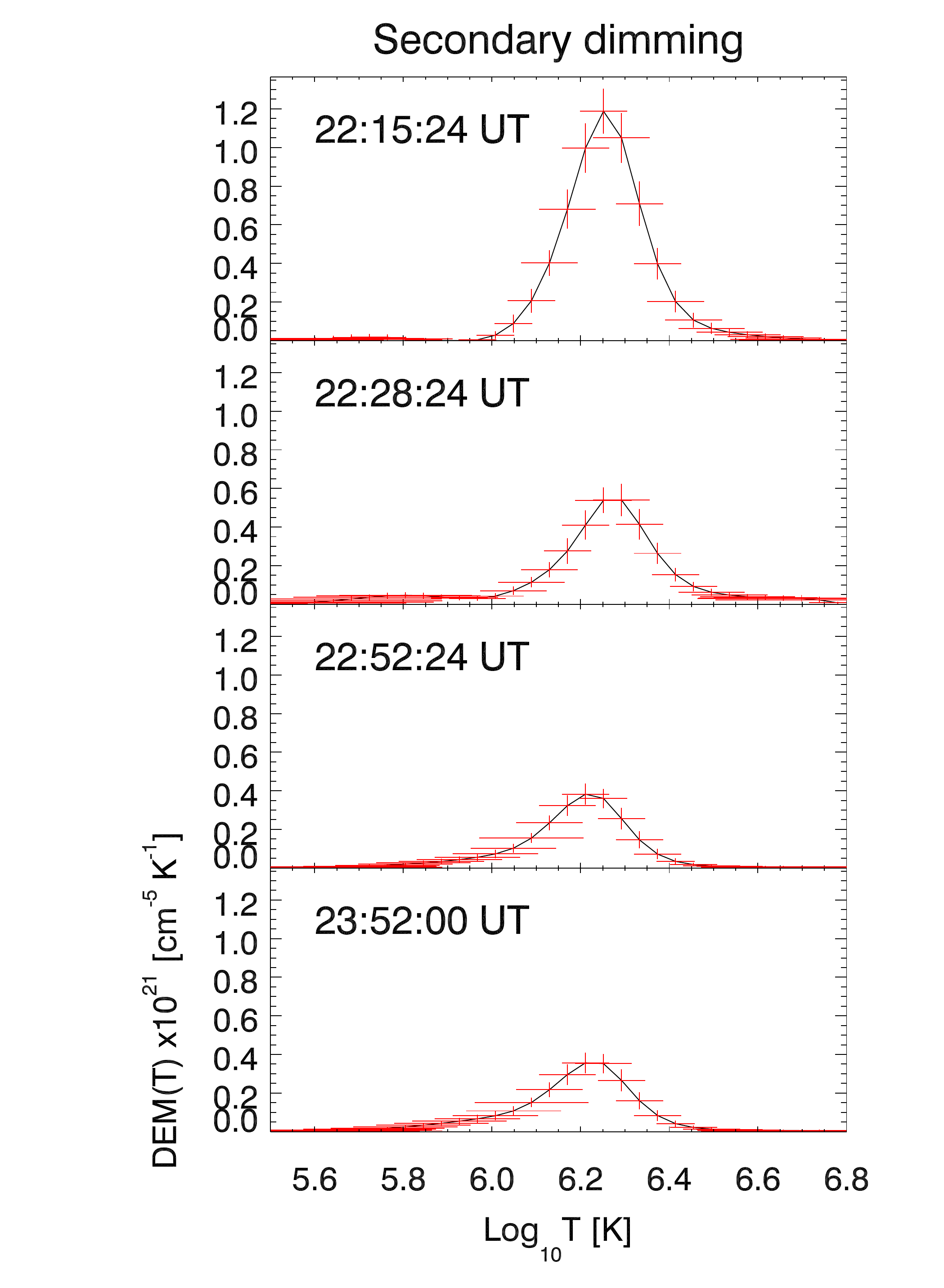}
\end{minipage}
\caption{DEM curves, for the event on 06 September 2011, derived from a subregion  in the core dimming region (left) and the secondary dimming region (right), as indicated by green and red boxes in Figure\,\ref{fig:plasma_evol_baseratio}, respectively.}
\label{fig:dem_evol}
\end{figure*}

For detailed analysis we selected subregions within the core and secondary dimming and checked the shape of the DEM curves. The evolution of these DEM profiles are shown in Figure\,\ref{fig:dem_evol}. Before the dimming region is formed, the DEMs in both the regions are quite high. The core dimming shows a double peak at 
$\log T[K]\,=\,6.20$ and $\log T[K]\,=\,6.45$, suggesting that plasma at quiet coronal temperatures as well as plasma at active region temperatures is included in the LOS of the selected core dimming region.  The secondary dimming shows one broad peak at $\log T[K]\,=\,6.25$, indicative that most plasma along the LOS is at typical quiet coronal temperatures.  As the dimming evolves, we see that the overall DEM curve diminishes drastically in both the regions. This indicates strong reduction of emission along the LOS due to depletion of plasma. In the core dimming region, it is noticeable, that the peak at $\log T[K]\,=\,6.45$ fully vanishes, indicating that active regions loops are ejected. In the secondary dimming region one can clearly see that not only the overall DEM distribution strongly reduces, but also that the peak moves to smaller temperatures, indicating plasma cooling associated with the density decrease (as is expected from adiabatic expansion). 

\begin{figure*}
\centering
\begin{minipage}[t]{0.4\textwidth}
\includegraphics[width=6.4cm]{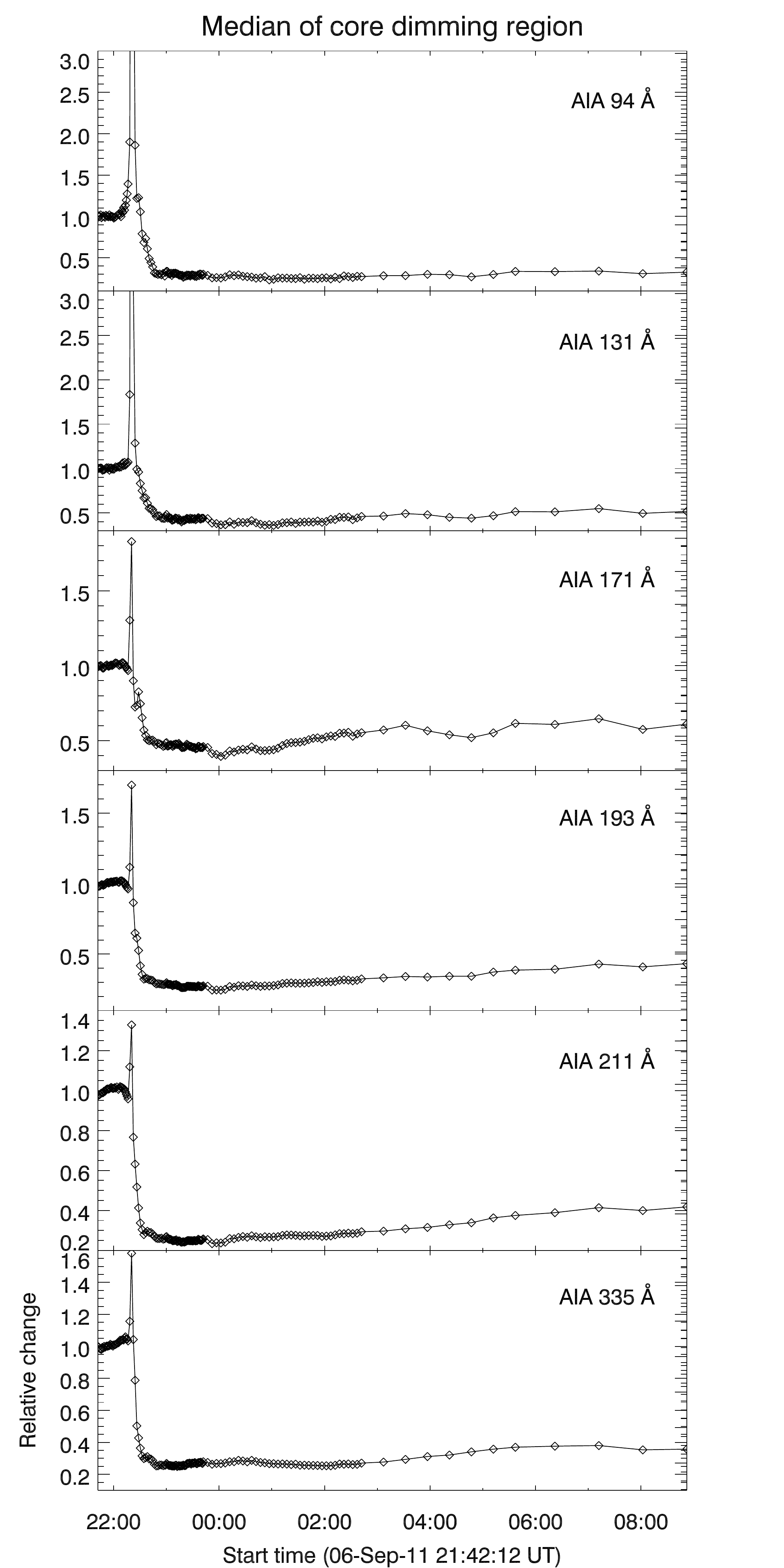}
\end{minipage}\hspace{-1cm}\begin{minipage}[t]{0.4\textwidth}~
\includegraphics[width=6.4cm]{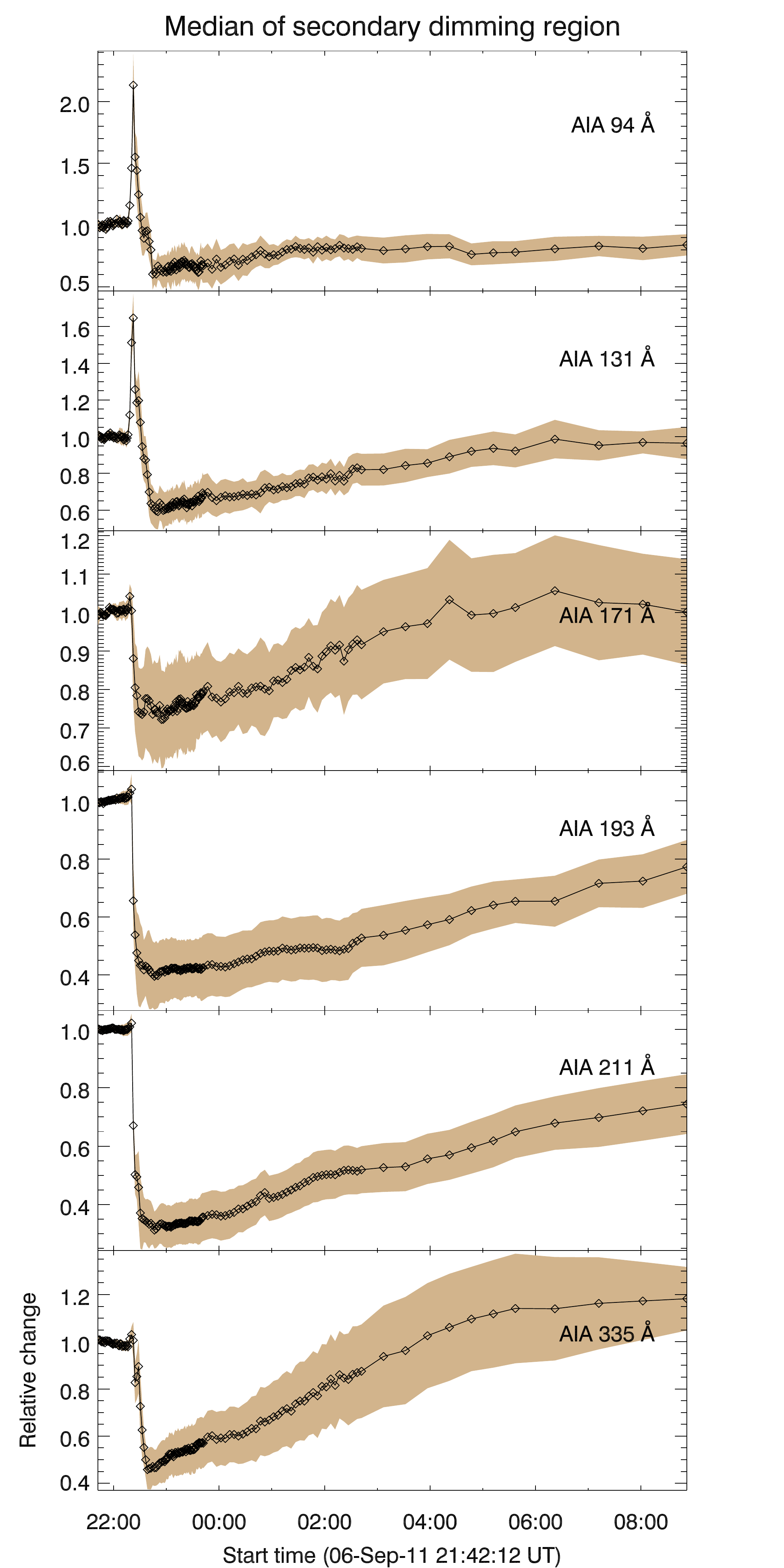}
\end{minipage}
\caption{Normalized lightcurves from the six SDO/AIA channels of the core dimming (left) and secondary dimming (right) for the event on 06 September 2011. Note that in the left panels the lightcurves are cut on the positive y-axis (dominated by the strong flare peak), to make the decrease due to the coronal dimming better visible. The region used for these lightcurves are indicated by green and red boxes  in Figure\,\ref{fig:plasma_evol_baseratio}, respectively.}
\label{fig:median_euv_lc}
\end{figure*}

To further study the evolution in the core and secondary dimming regions we derive lightcurves from the six EUV channels of AIA in the selected subregions. We study the evolution of the dimming region using the median of the lightcurves of pixels within the subregion. The selected regions of interest for core dimming (green) and secondary dimming (red) are marked in Figure\,\ref{fig:plasma_evol_baseratio}, left. The lightcurves, normalized to the pre-event value for each filter, are plotted in Figure\,\ref{fig:median_euv_lc} along with the absolute deviation. They give direct evidence on the percentage change of the emission with respect to the state before the dimming. We can see from Figure\,\ref{fig:median_euv_lc} that in the core dimming region, all AIA channels show a strong reduction in intensity (up to about 75\% decrease) immediately after the flare. We see the steepest changes within the first 10\,min after the flare start. It then takes 20 more minutes to reach a minimum. The lightcurves do not recover from the decrease for at least the next $\gtrsim$10\,hrs covered by our analysis. 

Within the secondary dimming region, the strongest decrease of $\approx$60--70\% is seen in the 193\,\AA, 211\,\AA\ and 335\,\AA\ channels. The other channels show a lesser decrease of $\approx$30--50\%. The secondary dimming region is not as impulsive as the core dimming region. The lightcurves are seen to decrease and reach a minimum after about 30\,min. However, unlike the core dimming region which stays at strongly reduced levels for the following $\gtrsim$10\,hrs, the intensity in the  secondary dimming begins to increase again within 10--30\,min after reaching its minimum.

\begin{figure*}
\centering
\begin{minipage}[t]{0.4\textwidth}
\includegraphics[width=6.4cm]{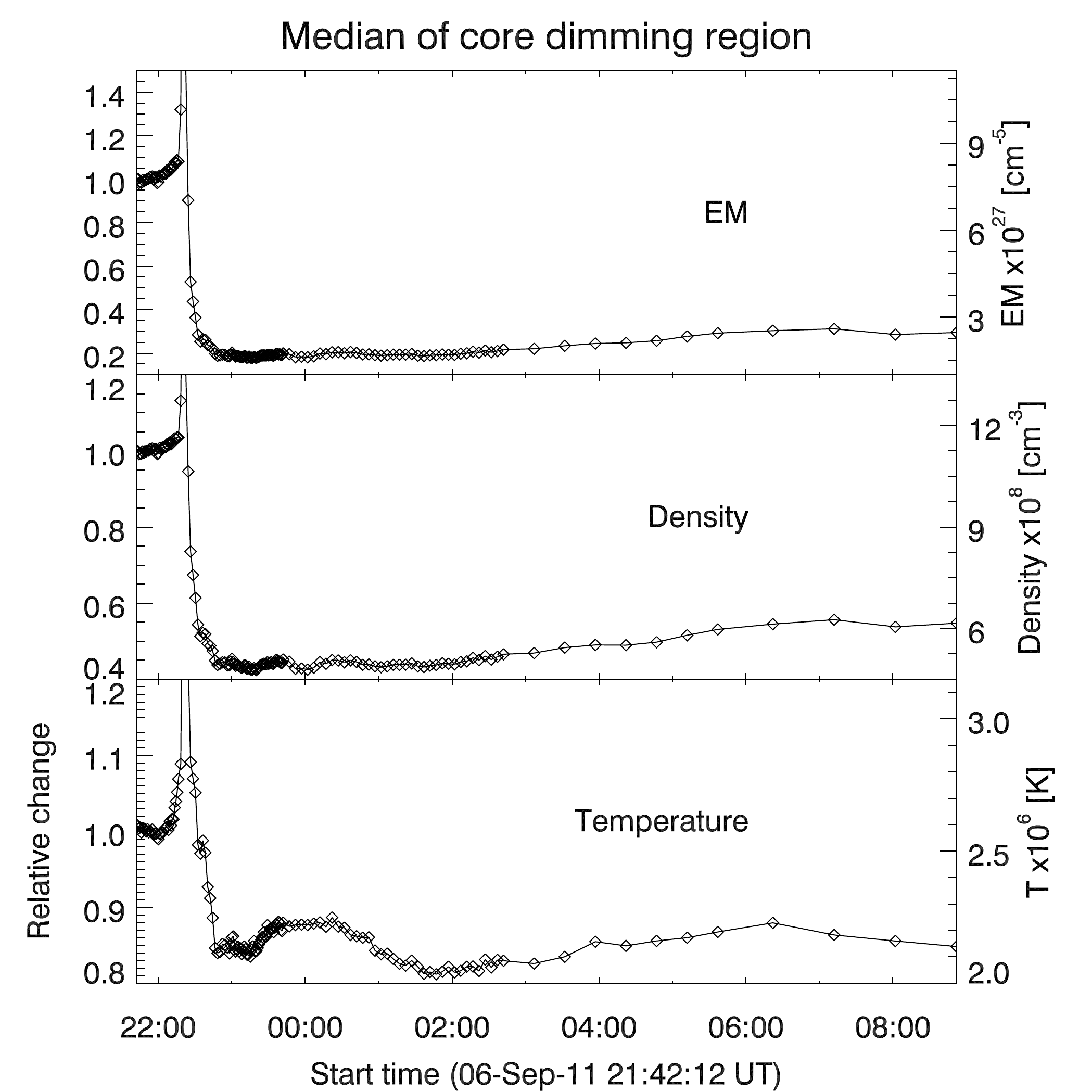}
\end{minipage}\hspace{-1cm}\begin{minipage}[t]{0.4\textwidth}~
\includegraphics[width=6.4cm]{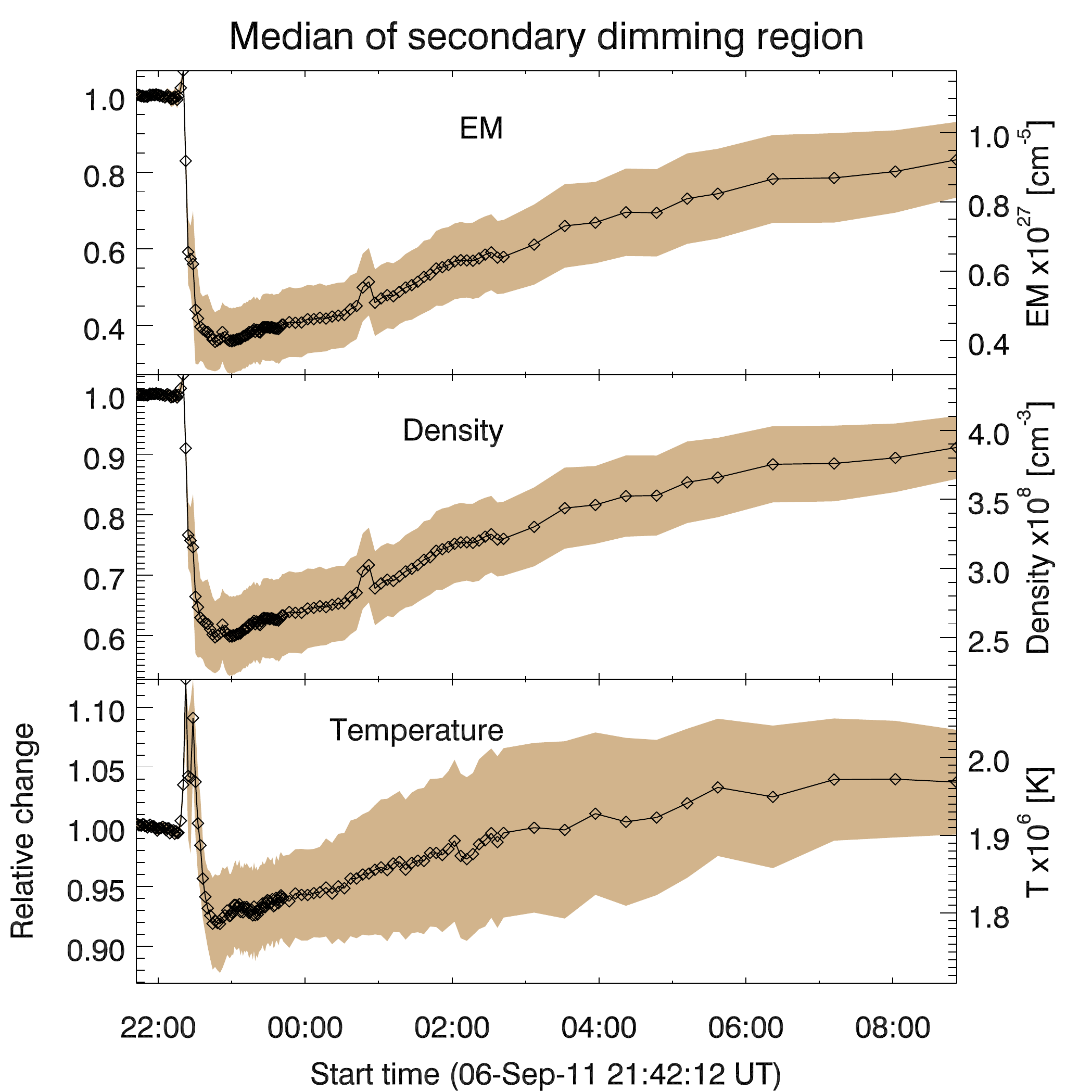}
\end{minipage} 
\caption{Time evolution of the plasma parameters within the core dimming (left) and secondary dimming (right) regions for the event on 06 September 2011. The left axes give the relative changes with respect to the pre-event, the right axes give the absolute values.}
\label{fig:median_plasma_lc}
\end{figure*}

We also studied the changes in the derived plasma parameters in a similar way and the results are shown in Figure\,\ref{fig:median_plasma_lc}. The EM and density in both the regions behave similar to the EUV lightcurves i.e. they drop abruptly after the flare in the dimming pixels. In the core dimming region there is a distinct density drop of 
58\% and temperature drop of 19\%. The secondary dimming region shows lesser decrease than the core dimming region, same as we saw in the AIA lightcurves. There is a 40\% decrease in density and a 8\% decrease in temperature. The larger decrease in density in both the core and secondary dimming regions as compared to the decrease in temperature suggests that the dimming regions are mainly formed by plasma evacuation rather than by temperature changes.

\pagebreak
\clearpage
\subsection{Event on 14 March 2012}
\label{ssec:14March2012}

The dimming event of 14 March 2012 was associated with an M2.8 class flare that occurred in NOAA Active Region 11432 (GOES peak at 15:08\,UT) and a CME with a speed of speed 411\,\kms, which is the slowest CME in our selection. It also had a moderate EUV wave of speed 485\,\kms\ associated with it.

\begin{figure*}[ht]
\centering
\includegraphics[width=\textwidth]{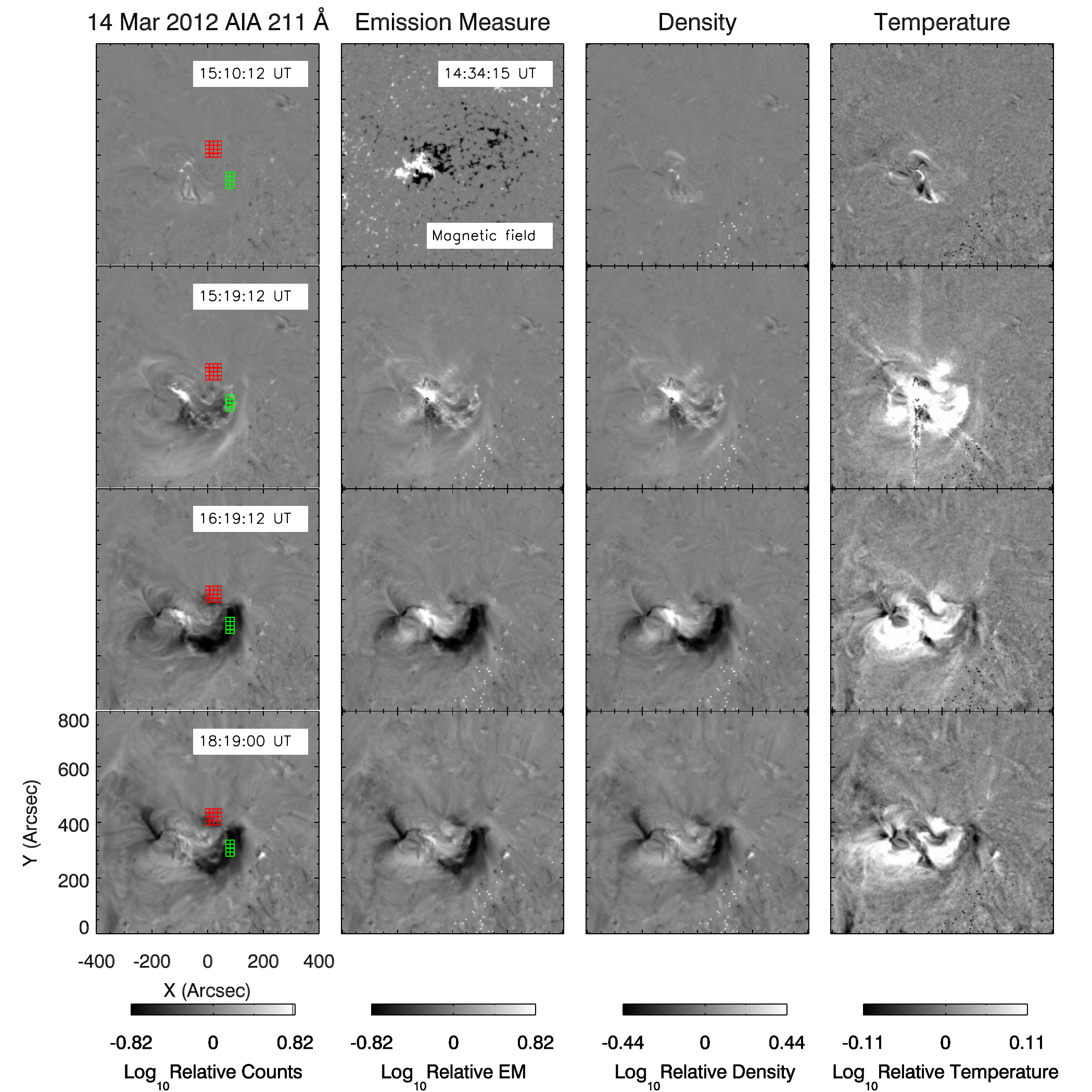}
\vspace{1cm}
\caption{Same as Figure\,\ref{fig:plasma_evol_baseratio} but for the event on 14 March 2012.}
\label{fig:2-plasma_evol_baseratio}
\end{figure*}

In Figure\,\ref{fig:2-plasma_evol_baseratio} we show the time evolution of the dimming in AIA 211\,\AA\ together with the EM, density and temperature maps. We see a clear coronal dimming formed, which is distinguishable from EM and density images but in the temperature images this region is partly obscured by the emission from the flare.

\begin{figure*}
\centering
\begin{minipage}[t]{0.4\textwidth}
\includegraphics[width=6cm]{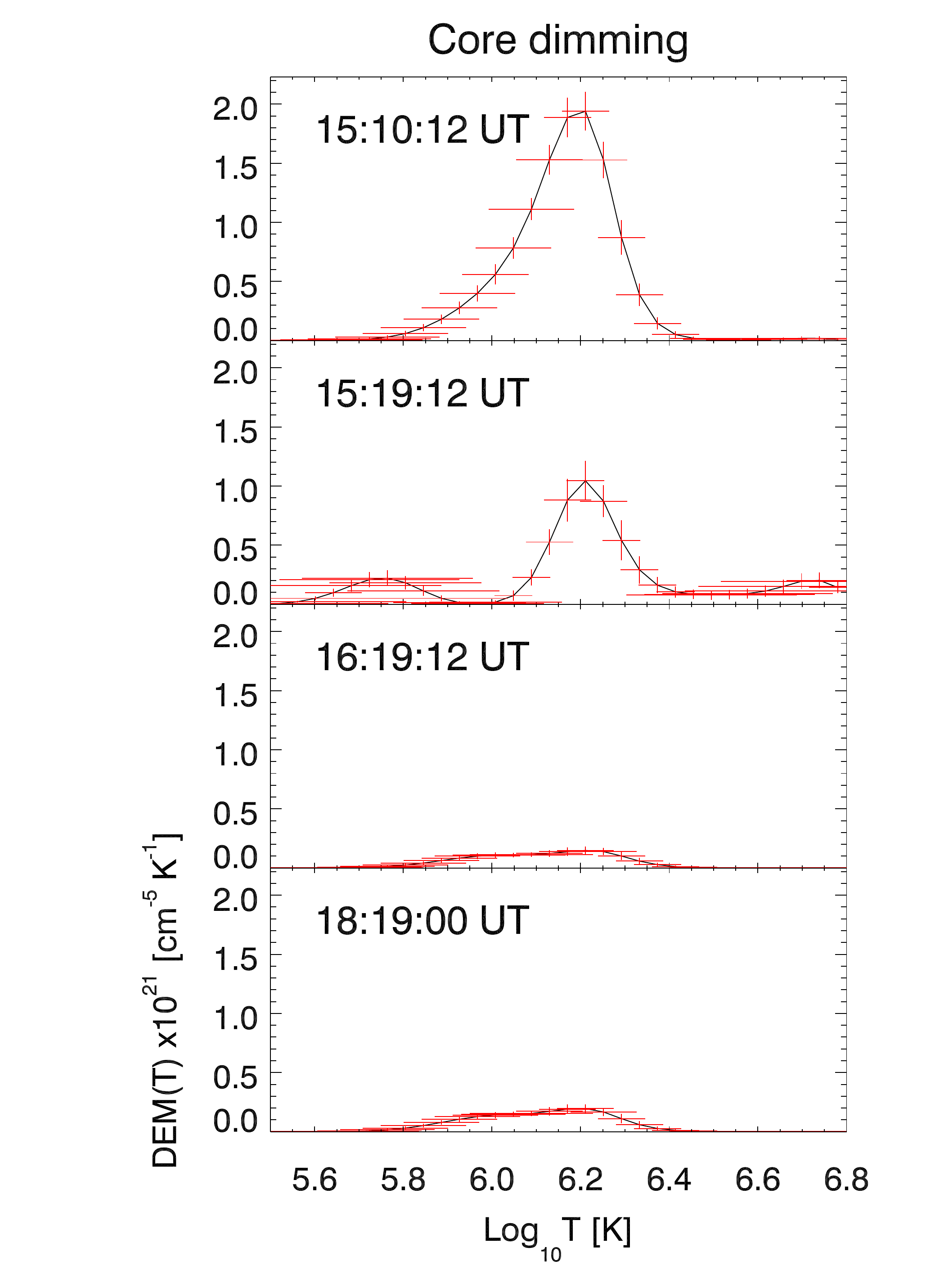}
\end{minipage}\hspace{-1cm}\begin{minipage}[t]{0.4\textwidth}~
\includegraphics[width=6cm]{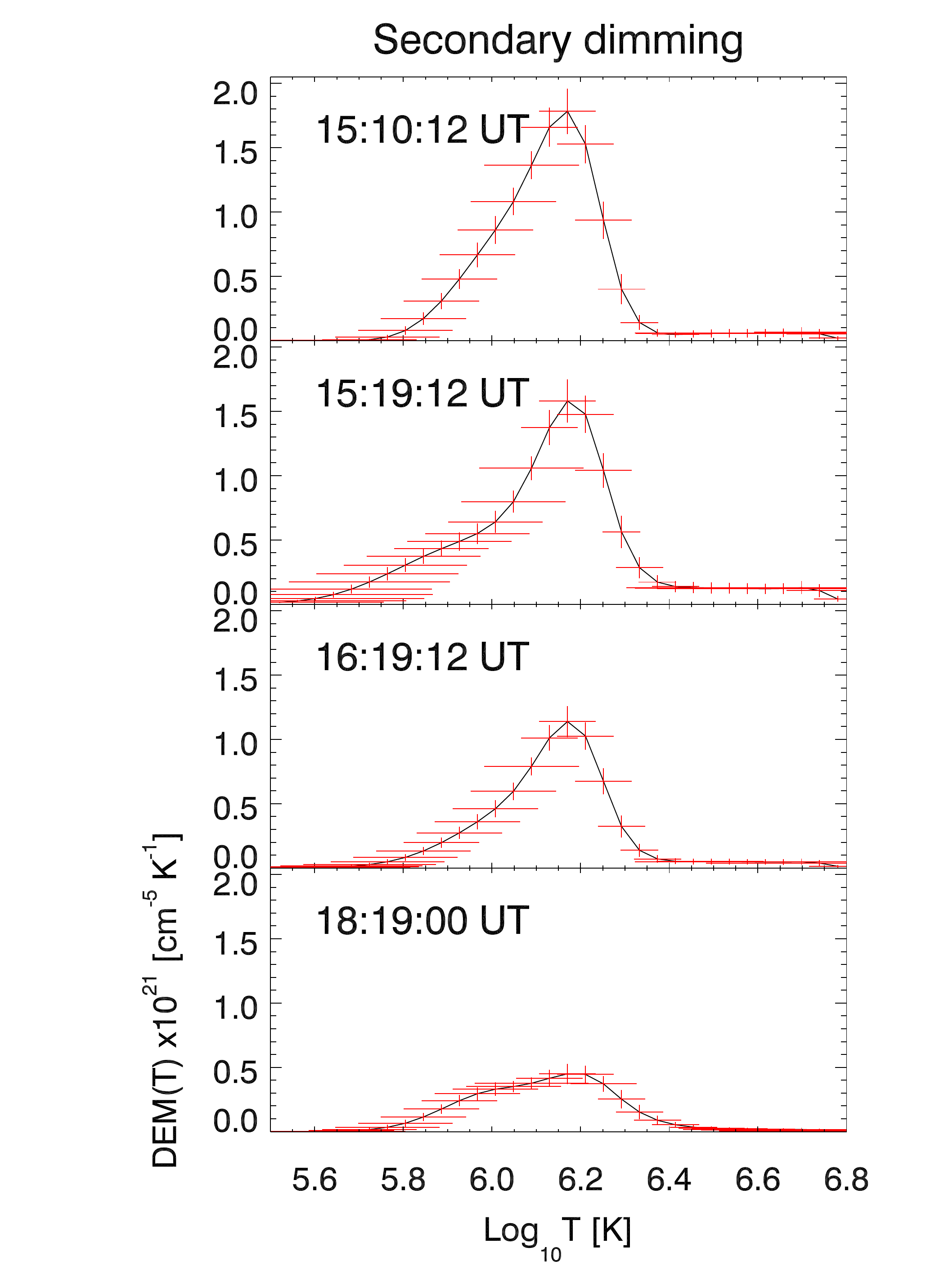}
\end{minipage}
\caption{Same as Figure\,\ref{fig:dem_evol} but for the event on 14 March 2012.}
\label{fig:2-dem_evol}
\end{figure*}

The DEM curves for this event are shown in Figure\,\ref{fig:2-dem_evol}. We see that the curves peak at around 
$\log T[K]\,=\,6.2$ in both the secondary and core dimming regions. As the dimming evolves, the DEM curves show a distinct decrease. In particular in the core dimming region, the overall DEM distribution drastically diminishes, indicating that a large portion of the emitting plasma along the LOS is removed. The pre-event peak values of the DEM curves 
($T = 1.7$ MK) indicate that most of the plasma that is removed is at  quiet coronal temperatures.

\begin{figure*}
\centering
\begin{minipage}[t]{0.4\textwidth}
\includegraphics[width=6.1cm]{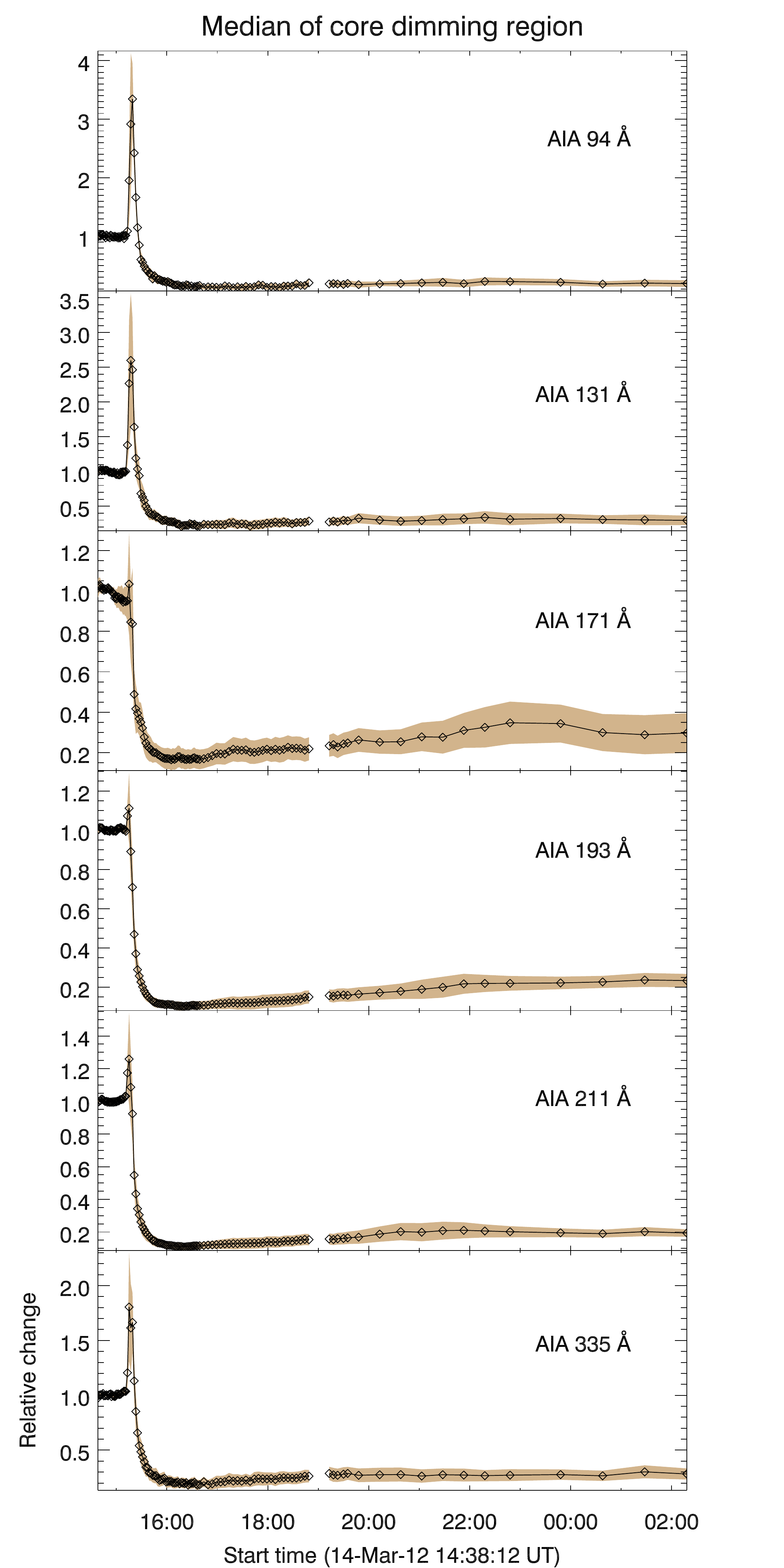}
\end{minipage}\hspace{-1cm}\begin{minipage}[t]{0.4\textwidth}~
\includegraphics[width=6.1cm]{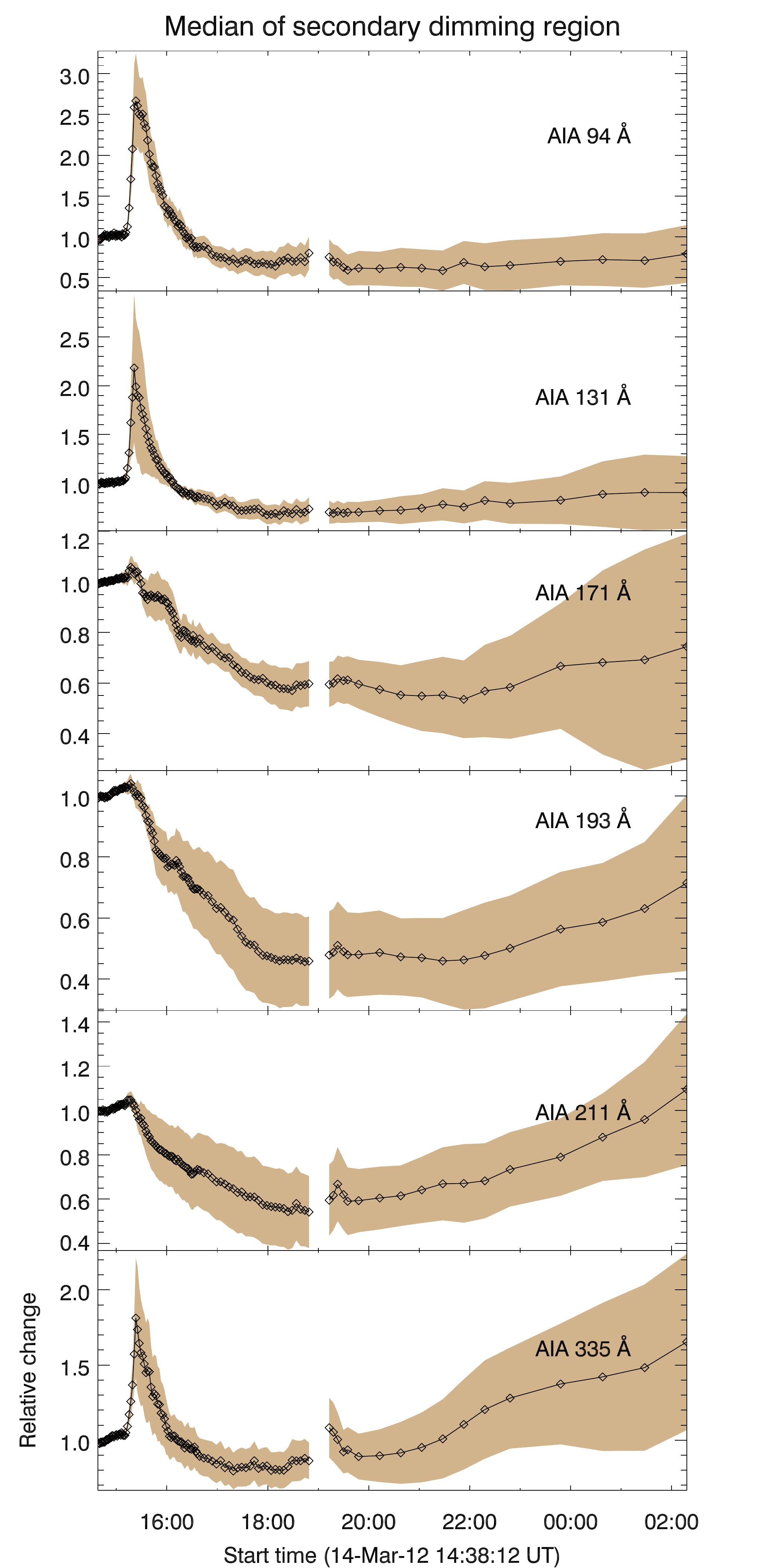}
\end{minipage}
\caption{Same as Figure\,\ref{fig:median_euv_lc} but for the event on 14 March 2012. We note there is a short data gap around 19 UT.}
\label{fig:2-median_euv_lc}
\end{figure*}

The AIA lightcurves in Figure\,\ref{fig:2-median_euv_lc} show a similar trend as in the event on 06 September 2011. In these lightcurves, we first see a peak due to the flare. In the core dimming region, we see immediately after this peak a decrease of more than 80\% in the lightcurves from all the AIA channels; the largest drop (90\%) is observed in the AIA 193\,\AA\ filter. This drastic decrease occurs within the first 30\,min after the flare start, and the lightcurves do not recover for at least the next 10\,hrs. The changes in the secondary dimming region evolve much slower than the core dimming region. The lightcurves reduce gradually, and take more than 2\,hrs to reach a minimum. Thereafter, they start to slowly rise again and return to background levels. For this region, the maximum intensity drop of 54\% is observed in the AIA 193\,\AA\ channel.

\begin{figure*}
\centering
\begin{minipage}[t]{0.4\textwidth}
\includegraphics[height=6.5cm]{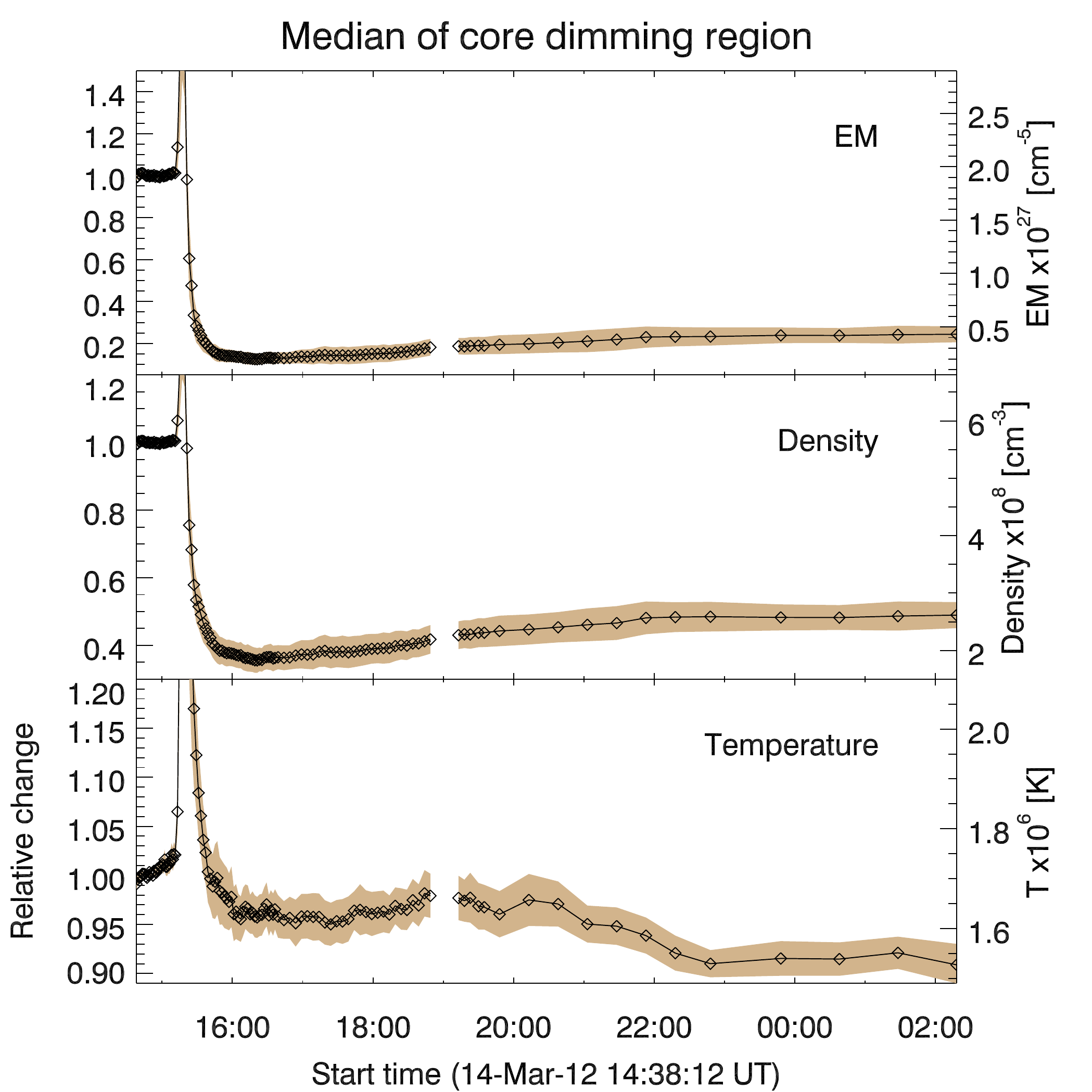}
\end{minipage}\hspace{-1cm}\begin{minipage}[t]{0.4\textwidth}~
\includegraphics[width=6.5cm]{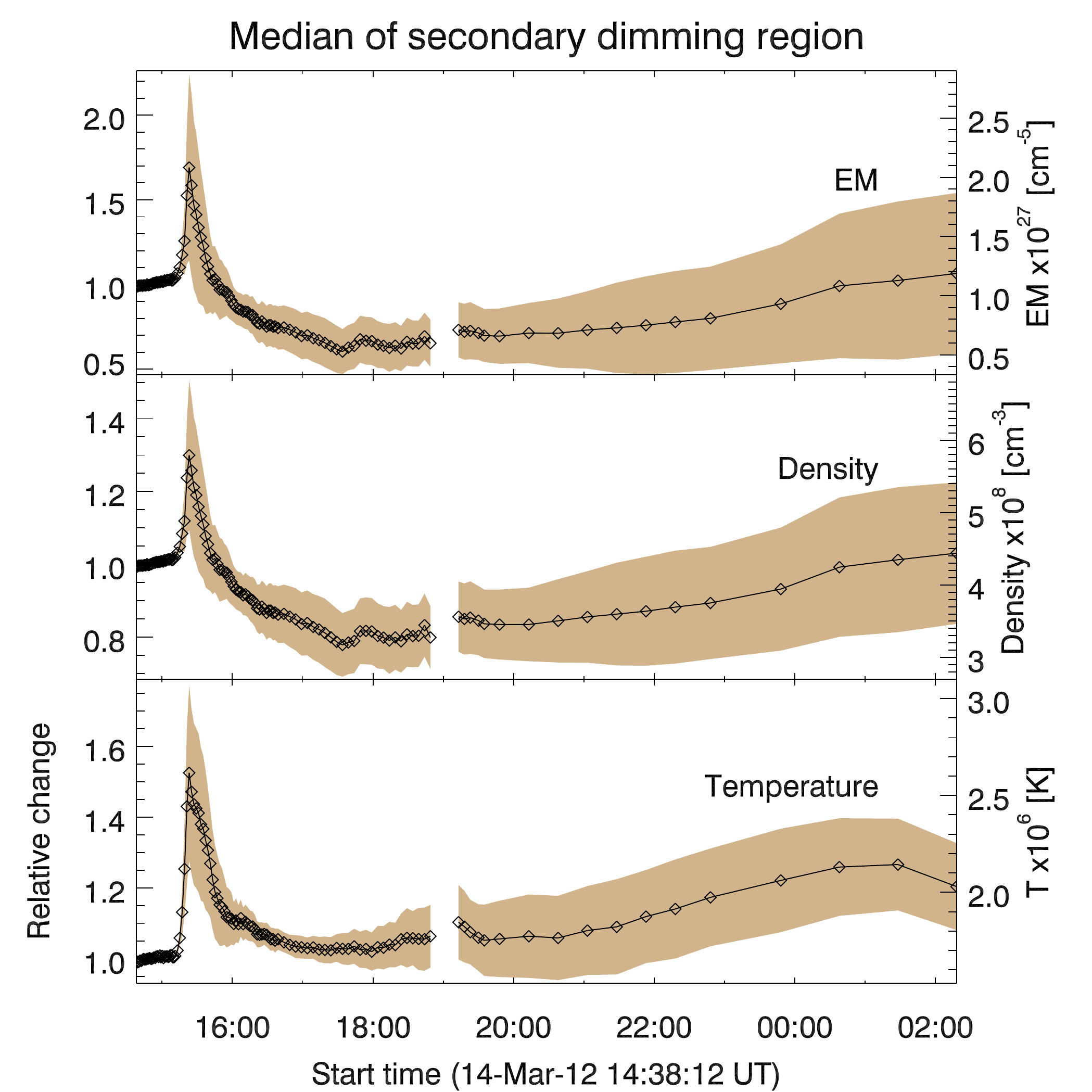}
\end{minipage}
\caption{Same as Figure\,\ref{fig:median_plasma_lc} but for the event on 14 March 2012.}
\label{fig:2-median_plasma_lc}
\end{figure*}

The time evolution of the plasma parameters is shown in Figure\,\ref{fig:2-median_plasma_lc}. 
In the core dimming region, we observe a steep gradient in the density and temperature evolution for about 20 min, where the density decreases by 64\% and the temperature by 9\%. There is very little variation in these values for the next 10\,hrs. Within the secondary dimming region, the plasma parameters take more than 2\,hrs to reach their minimum. The density decreases  by 22\%, and after the minimum is reached eventually starts to climb to reach background levels.  The temperature curve does not show any significant decrease from the background levels.  

\subsection{All other events}
\label{ssec:events_appendix}
Here we briefly discuss also all the other events we have studied (cf.\, Table\,\ref{table:events} ). The corresponding figures for these events are given in the Appendix. 

The dimming event on 01 August 2010 was associated to a small C3.2 class flare peaking at 07:24\,UT from NOAA Active Region 11092 and a CME with speed of 850\,\kms.  An EUV wave was also associated with it, with a speed of 312\,\kms.
Figure\,\ref{fig:3-plasma_evol_baseratio} shows images from the AIA 211\,\AA\ filter and the corresponding maps of the plasma parameters derived, illustrating the evolution of the dimming region. In Figure\,\ref{fig:3-median_euv_lc}, we show the AIA lightcurves, and in Figure\,\ref{fig:3-median_plasma_lc} the time evolution of the plasma parameters derived. The strongest density  decrease occurs within 10\,min after the flare start, followed by a more gradual further decrease, reaching a maximum drop of 68\% about 1 hr after flare start.  The temperature curve shows some fluctuation during the flare time, which we attribute to  uncertainties in the DEM reconstruction. Eventually it shows a steady decrease for about 2\,hrs, attaining a maximum drop of 25\%.  
In the secondary dimming region, the lightcurves evolve slowly and take about 40--60\,min to reach a minimum. The percentage decrease of density and temperature in the secondary dimming region is 31\% and 12\%, respectively.

The dimming event on 21 June 2011 was associated with a C7.7 class flare from NOAA Active Region 11236 (peak time 01:22\,UT) and a CME of speed 719\,\kms. No EUV wave was associated with the event.
Figure\,\ref{fig:4-plasma_evol_baseratio} shows the evolution of the dimming region. For this event the dimming region is initially obscured by loops along the LOS. This is characterized by the prolonged brightness in the AIA lightcurves and plasma parameters at the beginning of the event (see Figures\,\ref{fig:4-median_euv_lc} and \ref{fig:4-median_plasma_lc}). The loops finally open up to reveal a well formed dimming region. The changes in the core dimming region in this event are more gradual compared to the other events, which is due to the presence of loops. The density in the core dimming region reaches a minimum of 68\%, whereas the temperature shows only a 6\% decrease. In the secondary dimming region, the AIA lightcurves and the plasma parameters show many fluctuations that we attribute to the presence of loops along the LOS. The temperature curve does not show a clear decrease while the density curve reaches a minimum of 18\%.

The dimming event on 19 January 2012 was associated with an M3.2 class flare peaking at 13:44\,UT from NOAA Active Region 11402. This event was associated with the fastest CME within our event selection, with a speed 1120\,\kms\ but no EUV wave was associated.
The evolution of the dimming region is shown in Figure\,\ref{fig:5-plasma_evol_baseratio}. From the AIA lightcurves and time evolution of the plasma parameters, shown in Figures\,\ref{fig:5-median_euv_lc} and \ref{fig:5-median_plasma_lc} respectively, we see a pronounced dimming in both the core and secondary dimming regions. In the core dimming region, the fastest decrease occurs in the first 30\,min followed by a more gradual decrease thereafter, with the density dropping by 51\%. The secondary dimming reached a maximum drop in density by 
45\% in about 60\,min. In the secondary dimming region, the reconstructed temperature shows a more or less steady decrease, whereas in the core dimming region there are some fluctuations in the beginning.

The dimming event on 09 March 2012 was associated with an M6.3 class flare peaking at 03:22\,UT from NOAA Active Region 11429. This event was associated with a CME of speed 950\,\kms\ and an EUV wave of speed 690\,\kms.
Figure\,\ref{fig:6-plasma_evol_baseratio} shows maps of the AIA 211\,\AA\ filter and of the plasma parameters derived, illustrating the evolution of the dimming region. Lightcurves in the AIA channels and in the plasma parameters are shown in Figures\,\ref{fig:6-median_euv_lc} and \ref{fig:6-median_plasma_lc}, respectively. In this event, the plasma parameters show a rather gradual change. The density decreased by 52\%, reached after about 4\,hrs after the flare,  while the temperature  curve  shows almost no decrease. The evolution of the secondary dimming region is also slow and takes several hours to reach a minimum. The density decreases by 10\% but there is no clear decrease in the temperature curve.

In Tables\,\ref{tab:plasma_summary} and \ref{tab:euv_summary}, we 
summarize all the results obtained for the plasma parameters as well as for the changes in the AIA lightcurves for all six events under study.

\begin{table*}[htbp]
\centering
\begin{tabular}{| p{12mm} | p{8mm} | c | c | c | c | c | c | c | c | c | c | c |}
\hline
\multirow{2}{*}{Event} & GOES & \multicolumn{5}{c|}{Core dimming} & \multicolumn{5}{c|}{Secondary dimming} \\
\cline{3-12}
&flare & EM & Density & min. Density & Temp & min. Temp & EM & Density & min. Density & Temp & min. Temp\\
& class & \% & \% & $\times10^8$ cm$^{-3}$ & \% &  $\times10^6$ K & \% & \% &  $\times10^8$ cm$^{-3}$ & \% & $\times10^6$ K\\
\hline
\centering 01 Aug 2010 & C3.2 & 90 & 68 & 2.9 & 25 & 1.9 & 52 & 31 & 4.3 & 12 & 1.6 \\
\hline
\centering 21 Jun 2011 & C7.7 & 89 & 68 & 2.5 & 6 & 2.0 & 33 & 18 & 2.9 & - & -\\
\hline
\centering 06 Sep 2011 & X2.1 & 82 & 58 & 4.8 & 19 & 2.1 & 64 & 40 & 2.5 & 8 & 1.8 \\
\hline
\centering 19 Jan 2012 & M3.2 & 76 & 51 & 4.5 & 12 & 1.7 & 70 & 45 & 3.5 & 23 & 1.5 \\
\hline
\centering 09 Mar 2012 & M6.3 & 78 & 52 & 5.8 & - & - & 18 & 10 & 4.2 & - & - \\
\hline
\centering 14 Mar 2012 & M2.8 & 88 & 64 & 1.9 & 9 & 1.5 & 40 & 22 & 3.2 & - & -\\
\hline
\end{tabular}
\caption{The percentage decrease in the plasma parameters (EM, density, temperature) within the core and secondary dimming regions for all the events. Also given are the minimum values attained in density and temperature during the dimming evolution.
\label{tab:plasma_summary}}
\end{table*}

\section{Discussion}
\label{sec:sum_disc}


We have reconstructed DEM maps from SDO/AIA filtergrams for six CME-associated coronal dimming events to study the plasma characteristics and evolution in both the core and the secondary dimming regions. The most distinct difference we find is that in the core dimming region the plasma parameters show a much faster and deeper decrease than in the secondary dimming region, and stay at these low levels for the overall duration of our study.
In most cases, the core dimming region shows the steepest changes within 20--30\,min after the flare start,  while the secondary dimming evolution tends to be more gradual and takes longer to reach its minimum (30--90\,min). After reaching a minimum, within the core dimming region, there was not much change in the time evolution of the plasma parameters. They remained low and showed (almost) no signs of increase for the entire duration of this study ($\gtrsim$10\,hrs after the flare). This is consistent with the findings that CMEs may keep their connection to the Sun over days, as is  suggested from measuring bidirectional electron streams at 1AU \citep[e.g.,][]{Bothmer1996}.
On the other hand, in the secondary dimming region we observe a gradual increase  about 1--2\,hrs after the minimum was reached, indicative of replenishment of these coronal regions with plasma after the CME eruption. These findings suggest that the core dimming region corresponds to the footpoints of the erupting flux rope where the magnetic field lines open to interplanetary space enable continuous outflow of plasma, preventing the refill of this region, while the secondary dimming corresponds to overlying fields higher up in the corona that expand due to the CME eruption.  

In Table\,\ref{tab:plasma_summary} we have summarized the results of the plasma parameters for all the events. We have calculated the percentage decrease of the minimum value attained by the plasma parameters as compared to the pre-flare background for all the events. We have listed the percentage decrease in EM, density and temperature separately for the core dimming and the secondary dimming regions for all the events. In addition we also list the absolute minimum values attained by density and temperature for both these regions.  
We found that within the core dimming region the decrease in EM is 80--90\%, density decrease is 50--70\% and temperature decrease is 5--25\%. While within the secondary dimming we found the decrease in EM to be around 20--70\%, density decrease around 10--45\%; not all cases showed a decrease in temperature. 

From the results summarized in Table\,\ref{tab:plasma_summary} we see that for a given event the percentage decrease of each plasma parameter is significantly higher within the core dimming region than in the secondary dimming region. In some cases the changes within the core dimming region are almost twice more than the changes in the secondary dimming region. 
We also notice that for each event, the percentage decrease in temperature  is much lower than the decrease in EM and density, indicating that the main cause of the dimming is plasma evacuation rather than temperature changes. Comparing the maximum density and temperature drops observed, we find that in all cases the temperature decreases are smaller than would be predicted from adiabatic expansion, $T/T_0 = (n/n_0)^{\gamma-1}$ 
with $\gamma=5/3 $ for a fully ionized plasma 
\citep[e.g.][]{Aschwanden2004}.

However, at this stage, it is important to discuss the different limitations in the reconstructions of EM, density and temperature from the DEM analysis. In our analysis we have seen that some events do not reveal a clear temperature decrease or show strong fluctuations in the temperature evolution curves, even in cases where the reconstructed EM and density evolution showed a clear and smooth evolution. The estimates of EM and density (see Equations\,\ref{eq:em}, \ref{eq:density}) are determined from the integral over the whole DEM curve reconstructed, whereas the determination of the mean plasma temperature (see Equation\,\ref{eq:temp}) is strongly dependent on the shape of the DEM curve \citep{Cheng2012}. This explains why the reconstructed EM and density maps and curves are much more robust than the derived temperatures with regard to problems in the reconstructed DEM curves. Such problems may, e.g., arise due to uncertainties in the instrument response, due to saturation, blooming effects or scattered light from the bright flaring regions affecting the selected subegions in some filters at  some time steps. Or, they may also arise, when very different structures (e.g. flare and dimming) are included in the same selected pixel (note that we have binned our maps by $8\times8$ AIA pixels for better signal-to-noise ratio). In the selection of our subregions in the core and secondary dimmings  for detailed analysis, we tried to avoid areas that are affected by such effects, but we cannot assure this holds for all the AIA channels during all the time steps.  Thus, we caution on those curves that do not show a smooth evolution but strong fluctuations.

The effects discussed above are much less severe in the pre-event state, i.e. before the dimming and the associated strong flare emission take place. Comparing the pre-event densities and temperatures in the core and secondary dimming regions, which characterize the plasma that is later expanding or ejected from this region, we find the following. In the core dimming, the pre-event densities are in the range $6\times10^8$ to  $1.3\times10^9$ cm$^{-3}$ and the pre-event temperatures 1.7--2.6\,MK, whereas the corresponding values in the secondary dimming regions are   $4\times10^8$ cm$^{-3}$ to  $7\times10^8$ cm$^{-3}$ and 1.6--2.0\,MK, respectively. Systematically, in each event, both the pre-event densities and temperatures are smaller in the core than in the secondary dimming region.
 We also note that the event with the highest density and temperature (06 September 2011)  showed a double component in the pre-event DEMs   at $\log T[K]\,=\,6.20$ and $\log T[K]\,=\,6.45$ (cf. Figure\,\ref{fig:dem_evol}). The higher temperature component fully disappeared during the dimming, and the lower temperature component also strongly reduced. This together with the high density derived for the core dimming region, $1.1\times10^9$ cm$^{-3}$, implies that in this event also hot and dense active region loops were ejected from the selected region in the core dimming region.

\begin{table*}
\centering
\begin{tabular}{| c | c | c | c | c | c | c |}
\hline
\multirow{2}{*}{Event}  &  \multicolumn{6}{c|}{Core dimming / Secondary dimming (\%decrease)} \\
\cline{2-7}
~ & ~ 94 ~ & ~ 131 ~ & ~ 171 ~ & ~ 193 ~ & ~ 211 ~ & ~ 335\\
\hline
01 Aug 2010 ~ & ~ 83 / 41 ~ & ~ 79 / 27 ~ & ~ 82 / 31 ~ & ~ 81 / 51 ~ & ~ \colorbox{green}{84} / \colorbox{myred}{58} ~ & ~ \colorbox{green}{88} / \colorbox{myred}{55}  \\
\hline
21 Jun 2011 ~ & ~ 82 / 20 ~ & ~ 76 /  22 ~ & ~ 74 / \colorbox{myred}{53} ~ & ~ \colorbox{green}{91} / \colorbox{myred}{39} ~ & ~ \colorbox{green}{92} / 33 ~ & ~ 77 / 25 \\
\hline
06 Sep 2011 ~ & ~ \colorbox{green}{77} / 40 ~ & ~ 64 / 41 ~ & ~ 60 / 28 ~ & ~ \colorbox{green}{76} / \colorbox{myred}{61} ~ & ~ \colorbox{green}{77} / \colorbox{myred}{69} ~ & ~ 75/ 54 \\
\hline
19 Jan 2012 ~ & ~ 69 / 42 ~ & ~ 64 / 36 ~ & ~ 75 / 44 ~ & ~ \colorbox{green}{78} / 64 ~ & ~ \colorbox{green}{77} / \colorbox{myred}{77} ~ & ~ 67 / \colorbox{myred}{71} \\
\hline
09 Mar 2012 ~ & ~ 62 / 16 ~ & ~ 55 / 21 ~ & ~ 65 / \colorbox{myred}{48} ~ & ~ \colorbox{green}{80} / \colorbox{myred}{51} ~ & ~ \colorbox{green}{79} / 45 ~ & ~ 69 / 2 \\
\hline
14 Mar 2012 ~ & ~ \colorbox{green}{89} / 41 ~ & ~ 79 / 33 ~ & ~ 84 / \colorbox{myred}{46} ~ & ~ \colorbox{green}{90} / \colorbox{myred}{54} ~ & ~ \colorbox{green}{89} / \colorbox{myred}{46} ~ & ~  82 / 20 \\
\hline
\end{tabular}
\caption{The percentage decrease in emission in each of the AIA channels within the core and secondary dimming region for all the events. The largest and second largest values of decrease for each type of dimming are highlighted. Green corresponds to core and red to secondary dimming regions.\label{tab:euv_summary}
}
\end{table*}

In order see in which of the AIA filters we can observe the dimming most clearly, we also analyzed the time evolution of the dimming regions in the different AIA filters, and identified which one reveals the highest percentage decrease compared to the pre-event intensities. 
The results for all the events are shown in Table\,\ref{tab:euv_summary}. We found that within the core dimming region the highest decrease occurs in the 211\,\AA\ and 193\,\AA\ channels, with values of about 80 to 90\%. 
For the secondary dimming regions we found that again the 193\,\AA\ and 211\,\AA\ filter reveal the strongest decrease, with values of about 40 to 75\%, but for several events also the 171\,\AA\  filter sensitive to cooler plasma shows strong decreases up to 50\%.  
Other authors have also calculated percentage decrease in dimmings using the EIT 195\,\AA\ and 171\,\AA\ filters \citep{Chertok2005, McIntosh2007}.  They showed 40--50\% decrease in the 195\,\AA\ and up to 80\% in the 171\,\AA\ filter.

\section{Summary and Conclusions}
\label{sec:concl}

In this paper, we use SDO/AIA DEM analysis to study the plasma characteristics and  evolution of coronal dimmings caused by CMEs. The six events under study are associated with CMEs of speeds ranging from 400--1100\,\kms\ and flare classes from C3 to X2. We have constructed DEM maps and derived EM, density and temperature maps to study the time evolution of each dimming event over a duration of  12\,hrs. Particular focus was drawn on the differences between the core and the secondary dimming regions. Our main results and related conclusions are the following:
\begin{enumerate}
\item For each event, the decrease in density is much larger in the core dimming than in the secondary dimming region. In both types of dimmings, the decrease in density is consistently more pronounced than the decrease in temperature. In the core dimming region, we observe reductions in density of 50--70\% while the temperature drops by 5--25\%, which is smaller than is expected from adiabatic expansion.  In the secondary dimming region we observe density decreases of 10--45\%, but not all events reveal a measurable temperature decrease. 
\item In the core dimming region, the main changes in the plasma characteristics occur impulsively over a period  of $\approx$20--30 min after the start of the associated flare. The core dimmings occur in opposite polarity regions, localized in the active region, and remain evacuated for extended periods of time ($\gtrsim$10\,hrs). The secondary dimmings are more dispersed and  start to recover already after 1--2\,hrs indicating replenishment of the corona after the eruption.
\item In each event, the pre-event densities and temperatures are higher in the core than in the secondary dimming regions. The pre-event densities in the core dimming are in the range $n=6\times10^8$  to  $1.3\times10^9$ cm$^{-3}$ and the pre-event temperatures $T= 1.7 - 2.6$\,MK, whereas for the secondary dimming regions we obtain  $n=4\times10^8$ to  $7\times10^8$ cm$^{-3}$ and $T = 1.6-2.0$\,MK. These values and systematic differences indicate that the plasma in the core dimming region stems from active regions, whereas the consistently lower densities and temperatures in the secondary dimming regions are indicative of plasma from overlying regions higher in the corona. 
\item Both core and secondary dimming regions are best observed in the AIA 211\,\AA\ and 193\,\AA\ filters, with percentage decreases up to 80--90\% in the core dimming and 40--75\% in the secondary dimming regions with respect to the pre-event intensities.
\end{enumerate}
Our results confirm that coronal dimmings associated with CMEs are formed due to plasma evacuation rather than temperature changes, and thus 
coronal dimmings provide indeed a valuable alternative means to estimate the mass ejected by CMEs \citep[e.g.][]{Harrison2003,Mason2014,Aschwanden2016}.
In addition, we provide further support of the interpretation that the core dimming regions are formed at the location of the footpoints of the erupting flux rope. Our findings that the core dimming regions remain evacuated for $\gtrsim$10 hrs is indicative of magnetic field lines open to interplanetary space, which enable continuous outflow of plasma and preventing the refill of this region --- as it is expected for an erupting flux rope still connected to the Sun \citep[e.g.,][]{Temmer2017}. The secondary dimmings correspond to plasma from higher lying coronal regions that expand during the CME eruption, and start to be replenished already 1--2 hrs after the start of the event. 
Finally, we note that the gradual evolution and replenishment derived for the secondary dimming regions may also bear general implications and characteristic time scales for the formation of the corona.





\acknowledgments

This work is supported by the Austrian Space Applications Program of the Austrian Research Promotion Agency FFG (ASAP-11 4900217) and the Austrian Science Fund (FWF): P24092-N16.

%




\clearpage
\appendix
\section{}
Here we show the figures for all the events discussed in Section\,\ref{ssec:events_appendix}
\subsection{Event on 01 August 2010}
\label{ssec:01Aug2010}

\begin{figure*}[ht]
\centering
\includegraphics[width=18.6cm]{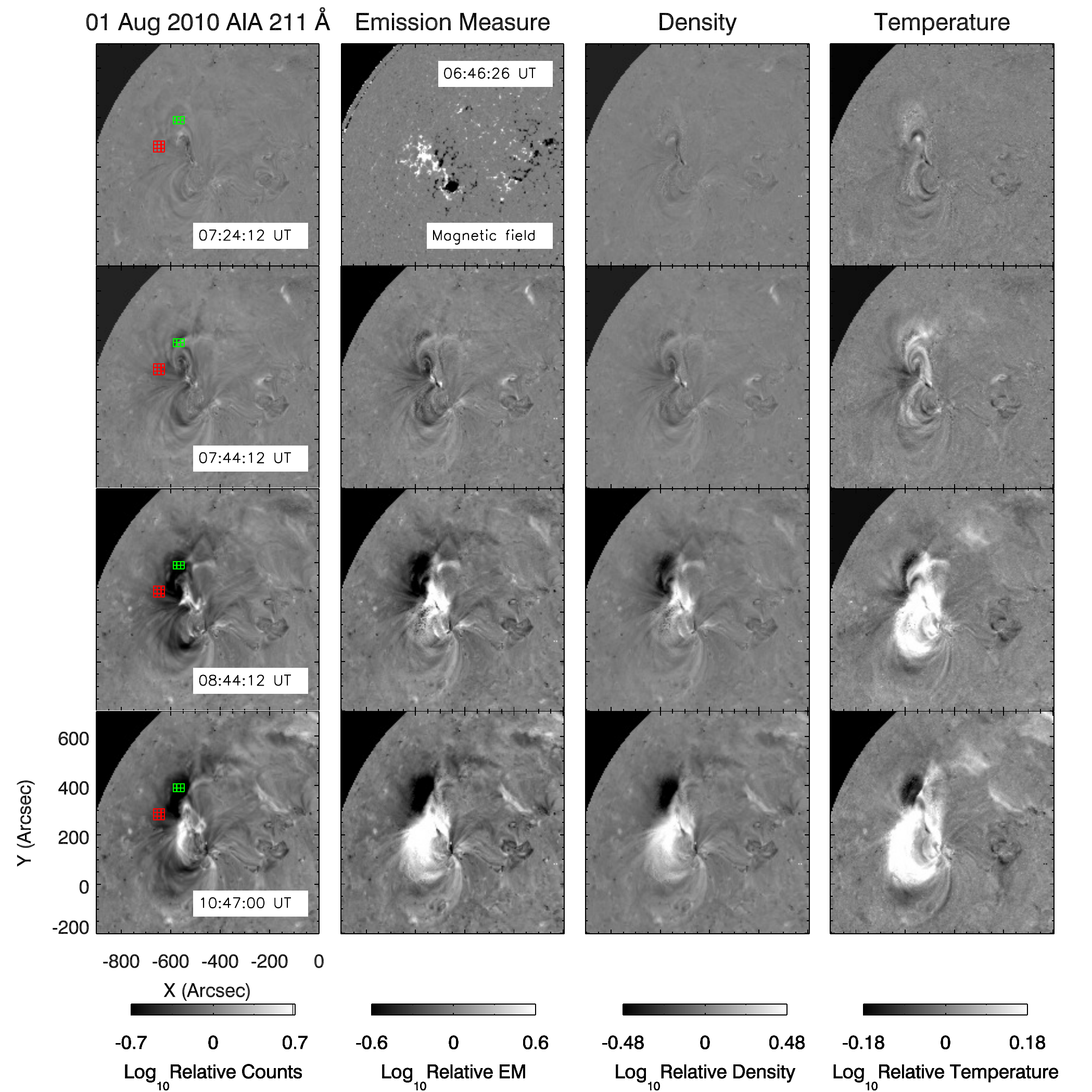}
\caption{Same as Figure\,\ref{fig:plasma_evol_baseratio} but for the event on 01 August 2010.}
\label{fig:3-plasma_evol_baseratio}
\end{figure*}


\begin{figure}
\centering
\begin{minipage}[t]{0.4\columnwidth}
\includegraphics[width=6.1cm]{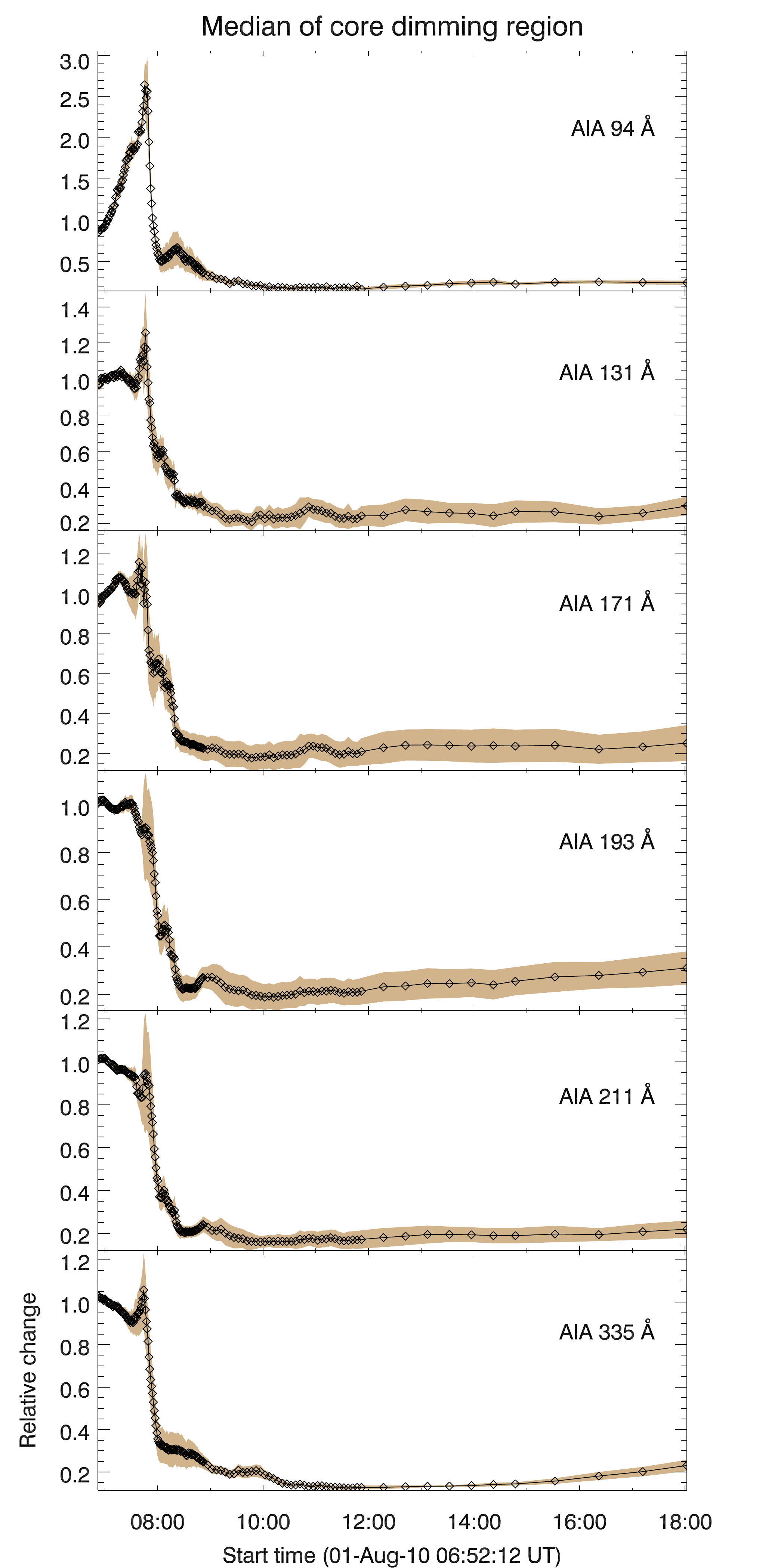}
\end{minipage}\hspace{-1cm}\begin{minipage}[t]{0.4\columnwidth}~
\includegraphics[width=6.1cm]{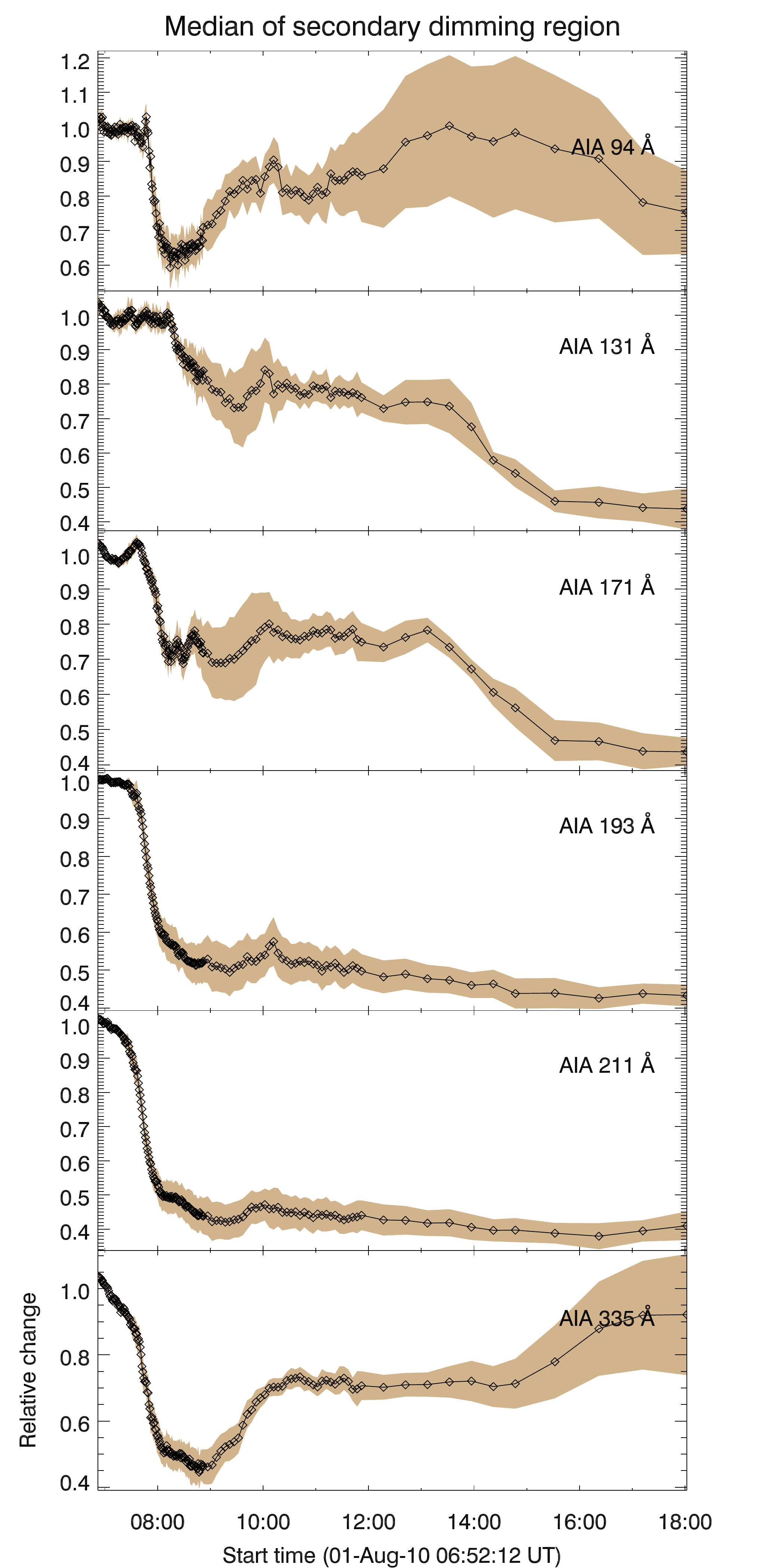}
\end{minipage}
\caption{Same as Figure\,\ref{fig:median_euv_lc} but for the event on 01 August 2010.}
\label{fig:3-median_euv_lc}
\end{figure}

\begin{figure*}
\centering
\begin{minipage}[t]{0.4\columnwidth}
\includegraphics[width=6.5cm]{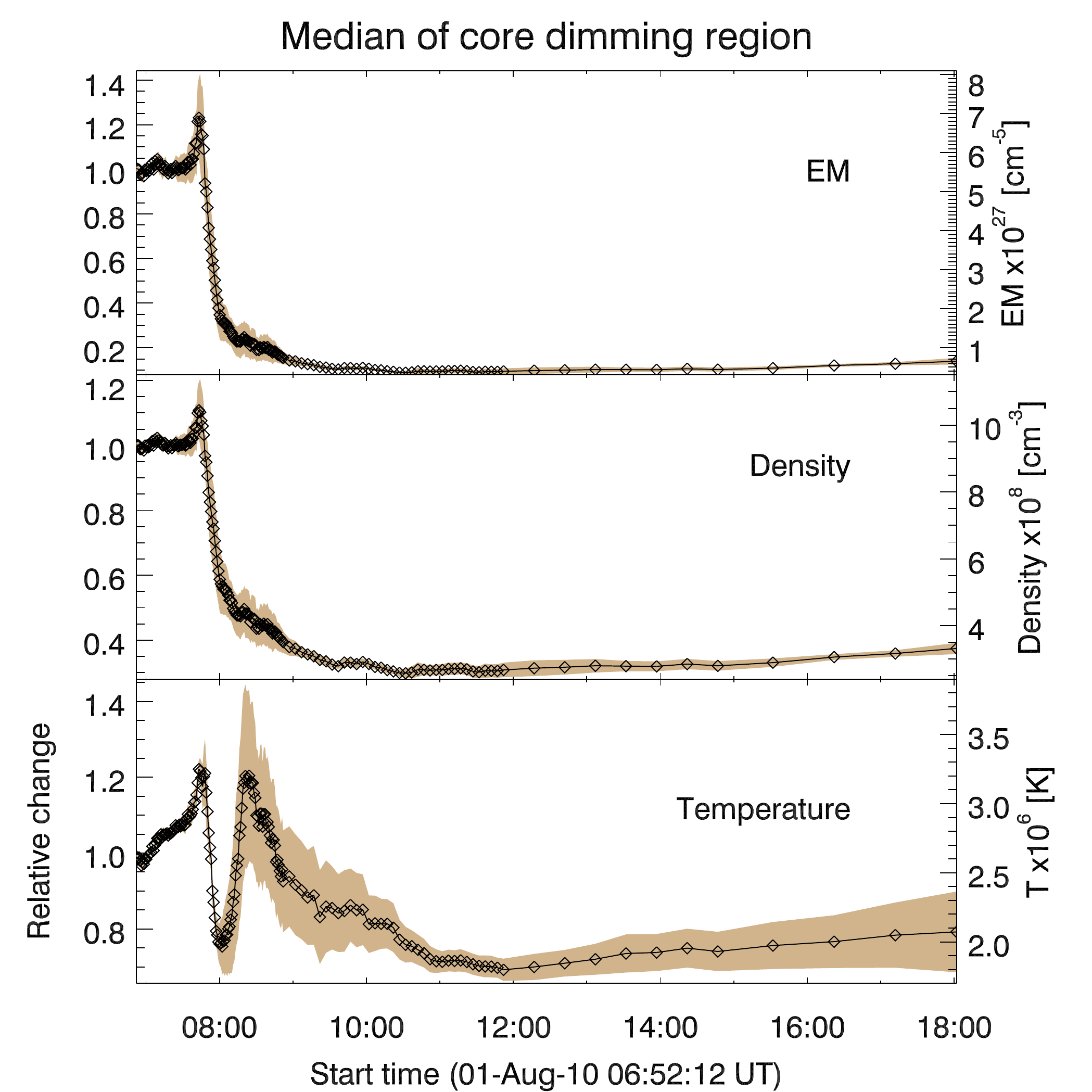}
\end{minipage}\hspace{-1cm}\begin{minipage}[t]{0.4\columnwidth}~
\includegraphics[width=6.5cm]{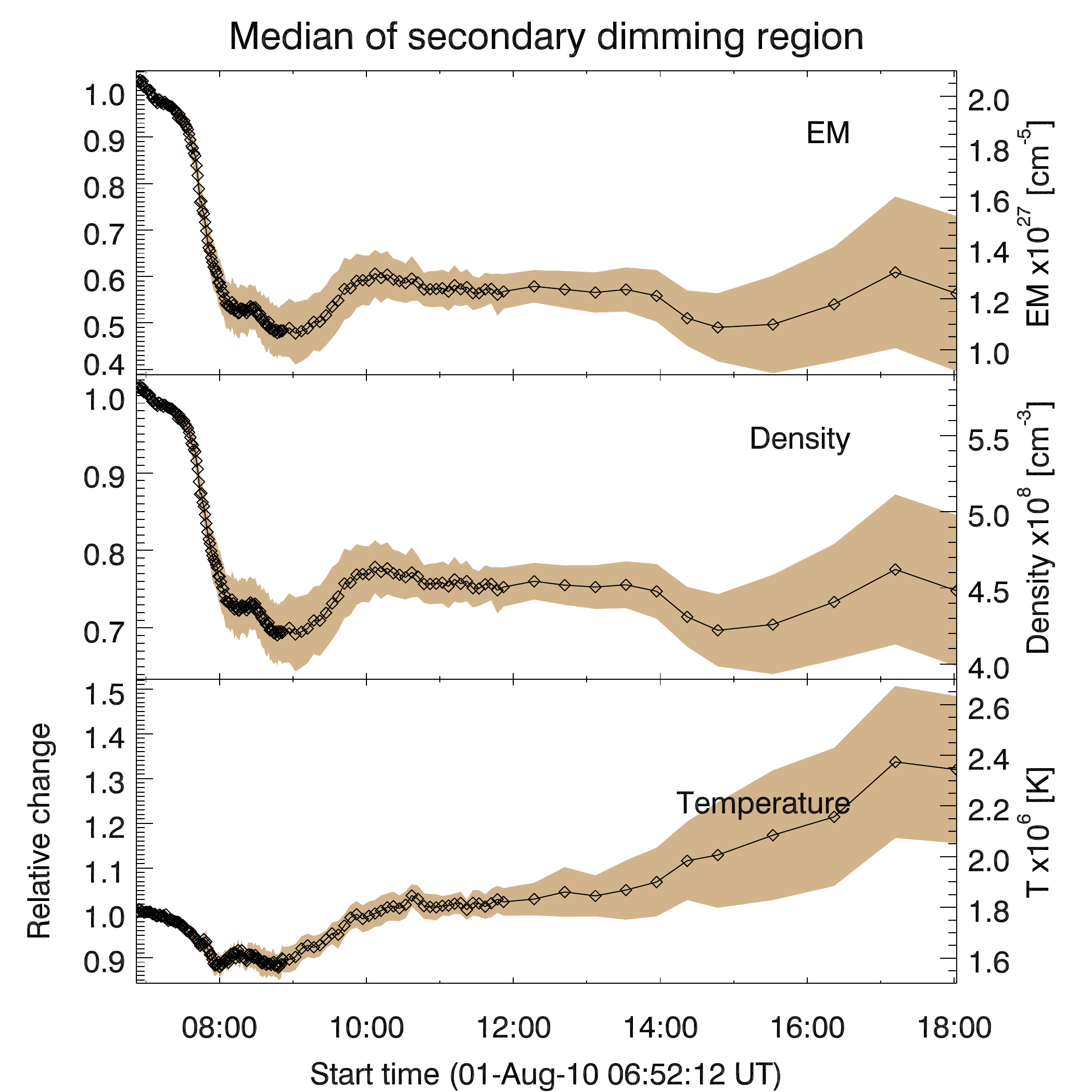}
\end{minipage}
\caption{Same as Figure\,\ref{fig:median_plasma_lc} but for the event on 01 August 2010.}
\label{fig:3-median_plasma_lc}
\end{figure*}

\clearpage

\subsection{Event on 21 June 2011}
\label{ssec:21Jun2011}

\begin{figure*}[ht]
\centering
\includegraphics[width=\textwidth]{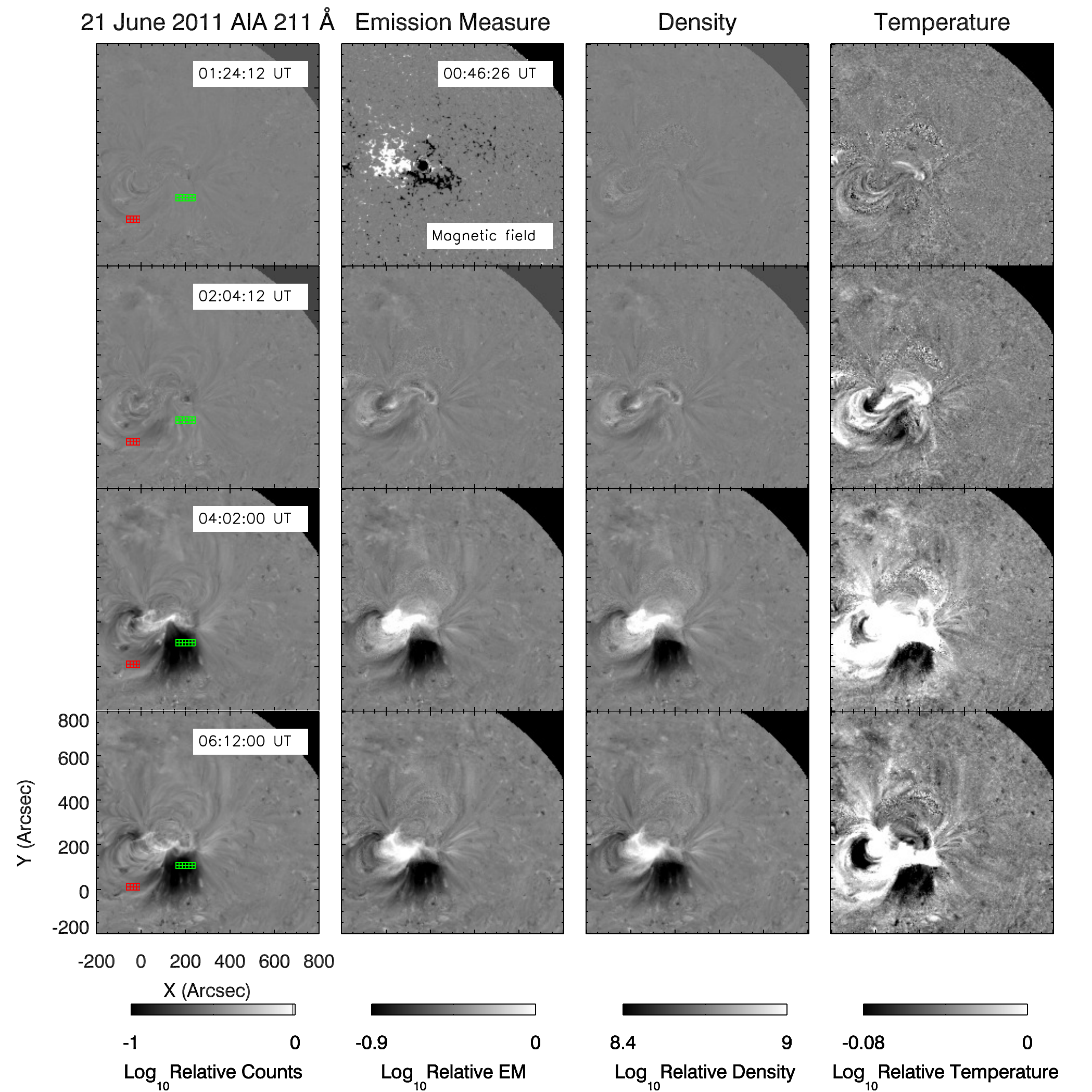}
\vspace{1cm}
\caption{Same as Figure\,\ref{fig:plasma_evol_baseratio} but for the event on 21 June 2011.}
\label{fig:4-plasma_evol_baseratio}
\end{figure*}


\begin{figure*}
\centering
\begin{minipage}[t]{0.4\textwidth}
\includegraphics[width=6.1cm]{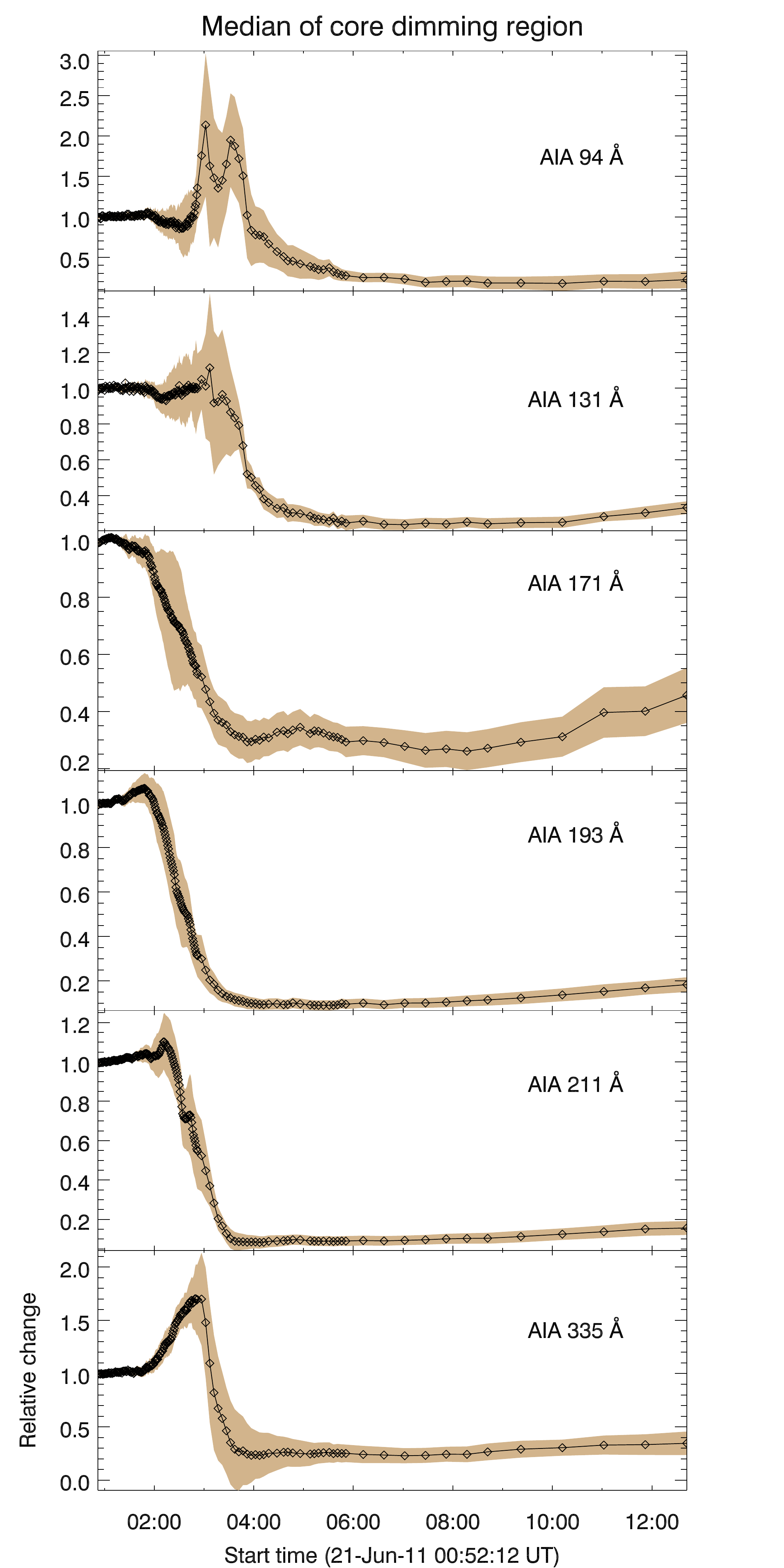}
\end{minipage}\hspace{-1cm}\begin{minipage}[t]{0.4\textwidth}~
\includegraphics[width=6.1cm]{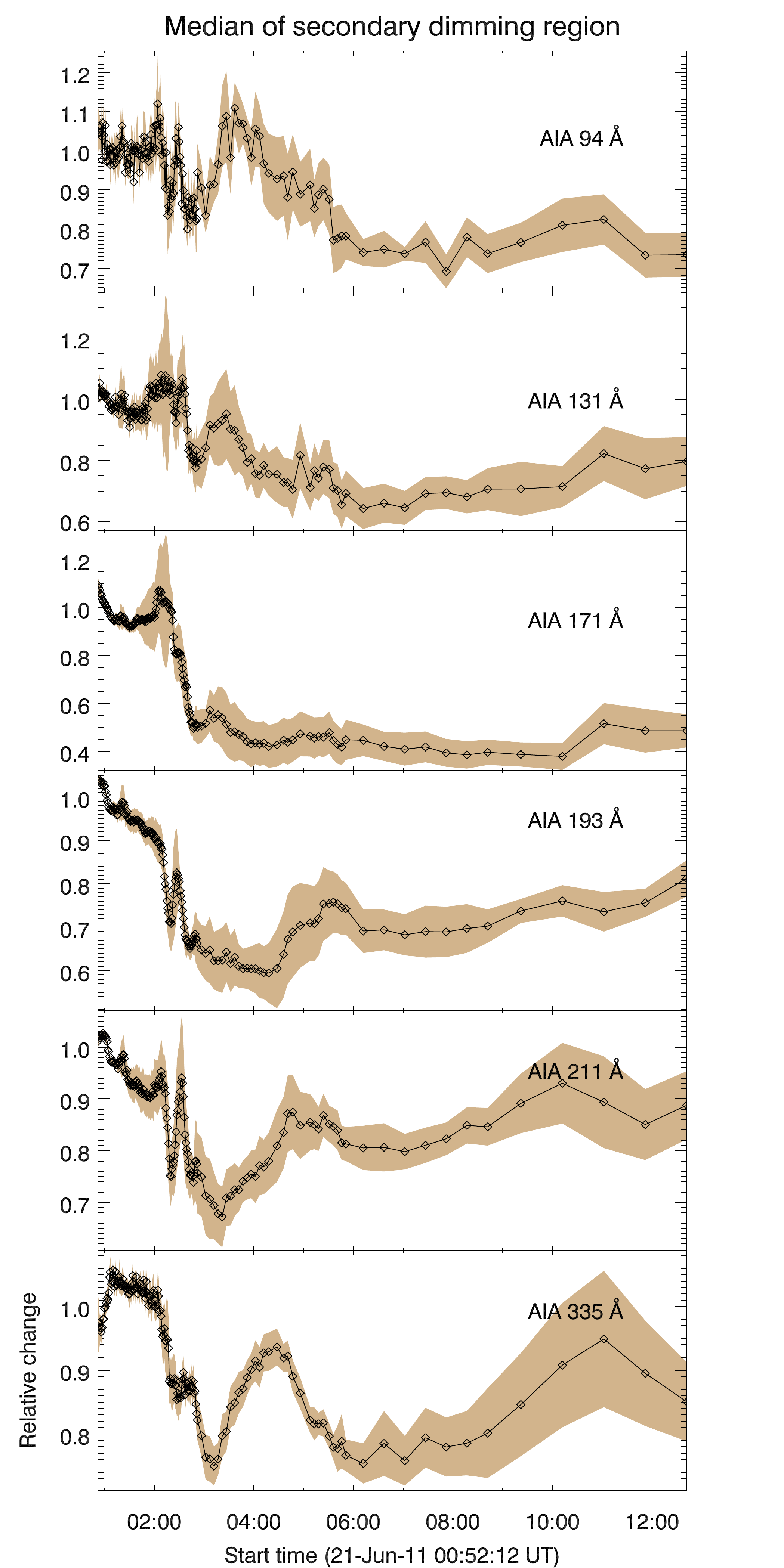}
\end{minipage}
\caption{Same as Figure\,\ref{fig:median_euv_lc} but for the event on 21 June 2011.}
\label{fig:4-median_euv_lc}
\end{figure*}

\begin{figure*}
\centering
\begin{minipage}[t]{0.4\textwidth}
\includegraphics[width=6.5cm]{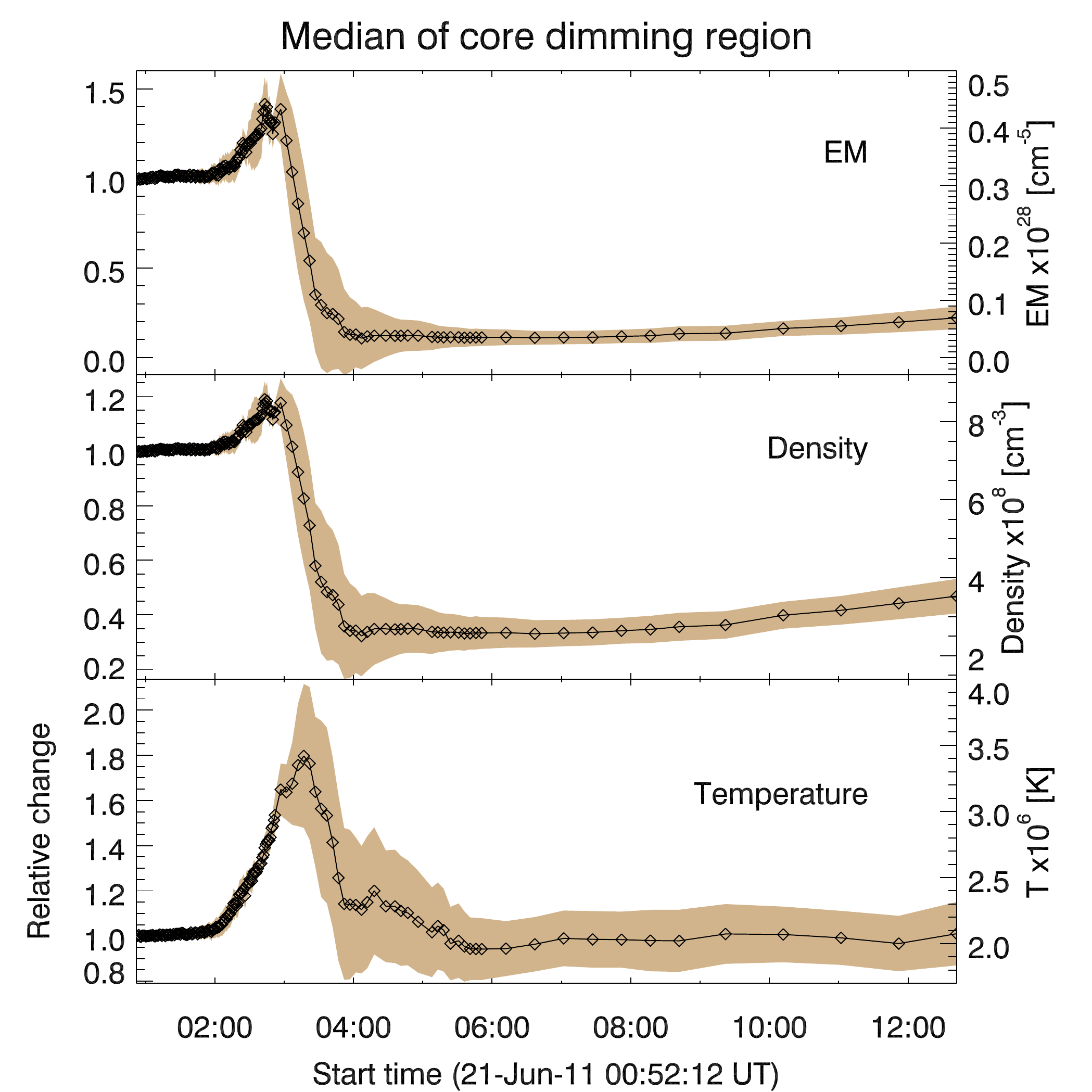}
\end{minipage}\hspace{-1cm}\begin{minipage}[t]{0.4\textwidth}~
\includegraphics[width=6.5cm]{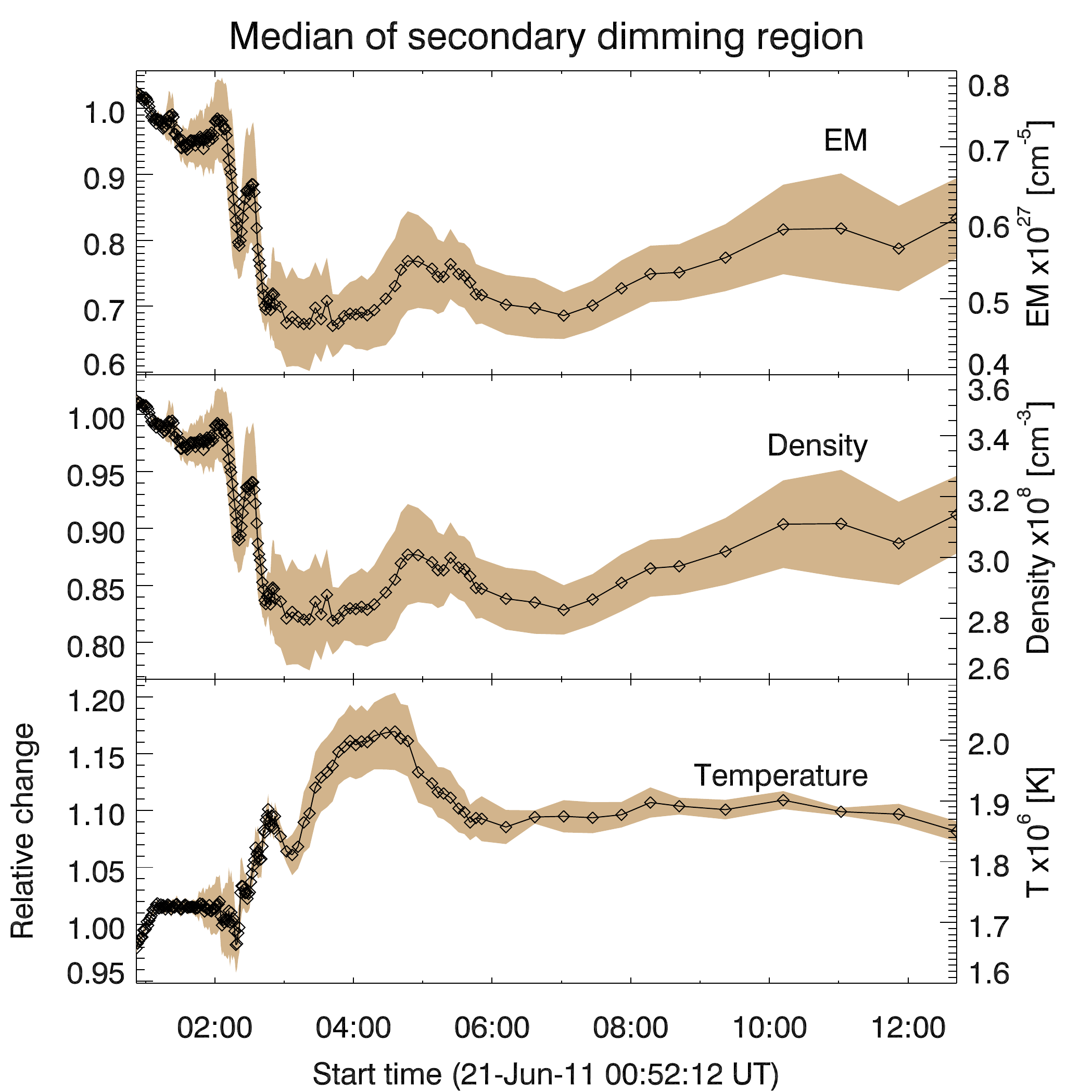}
\end{minipage}
\caption{Same as Figure\,\ref{fig:median_plasma_lc} but for the event on 21 June 2011.}
\label{fig:4-median_plasma_lc}
\end{figure*}

\clearpage

\subsection{Event on 19 January 2012}
\label{ssec:19Jan2012}

\begin{figure*}[ht]
\centering
\includegraphics[width=\textwidth]{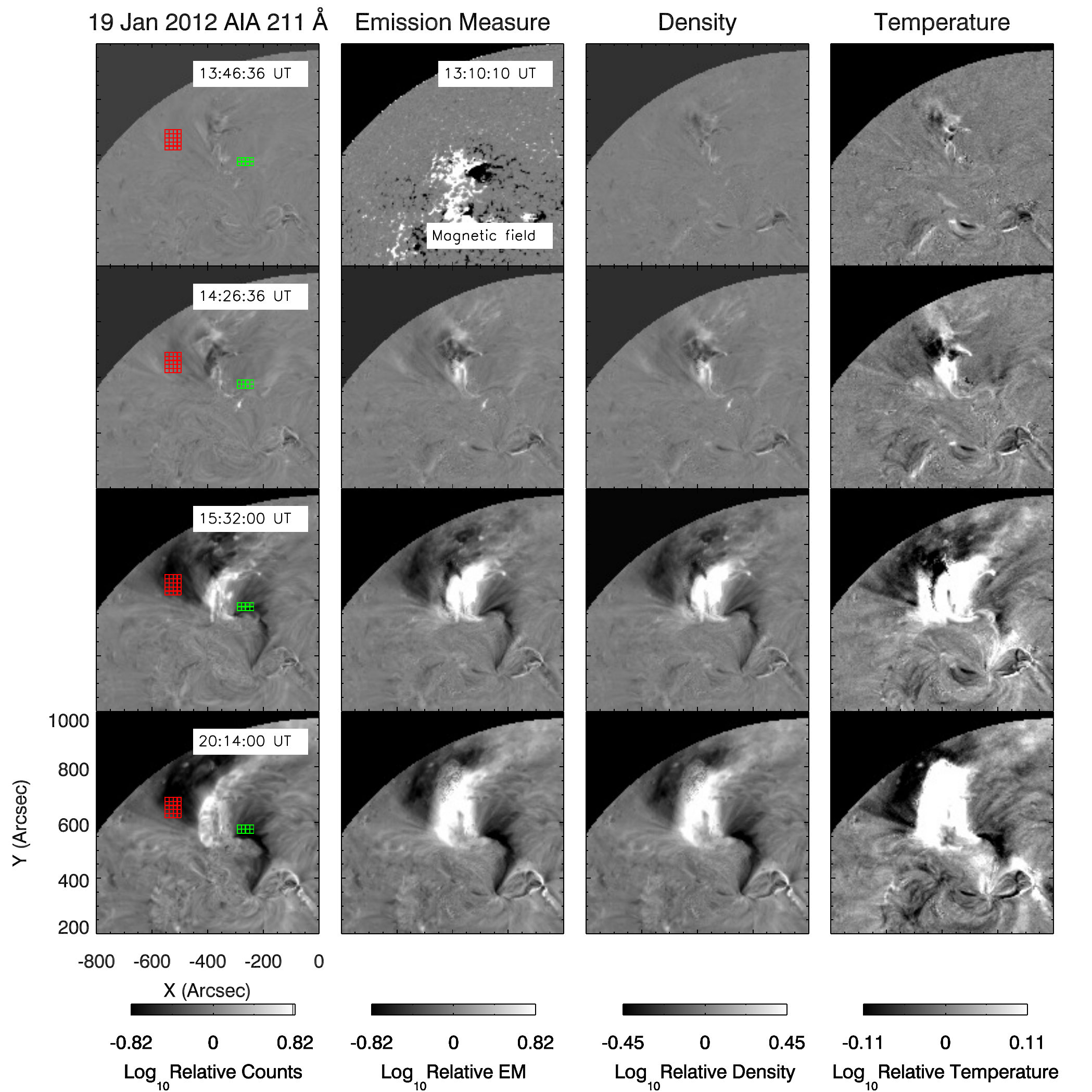}
\vspace{1cm}
\caption{Same as Figure\,\ref{fig:plasma_evol_baseratio} but for the event on 19 January 2012.}
\label{fig:5-plasma_evol_baseratio}
\end{figure*}


\begin{figure}
\centering
\begin{minipage}[t]{0.4\columnwidth}
\includegraphics[width=6.1cm]{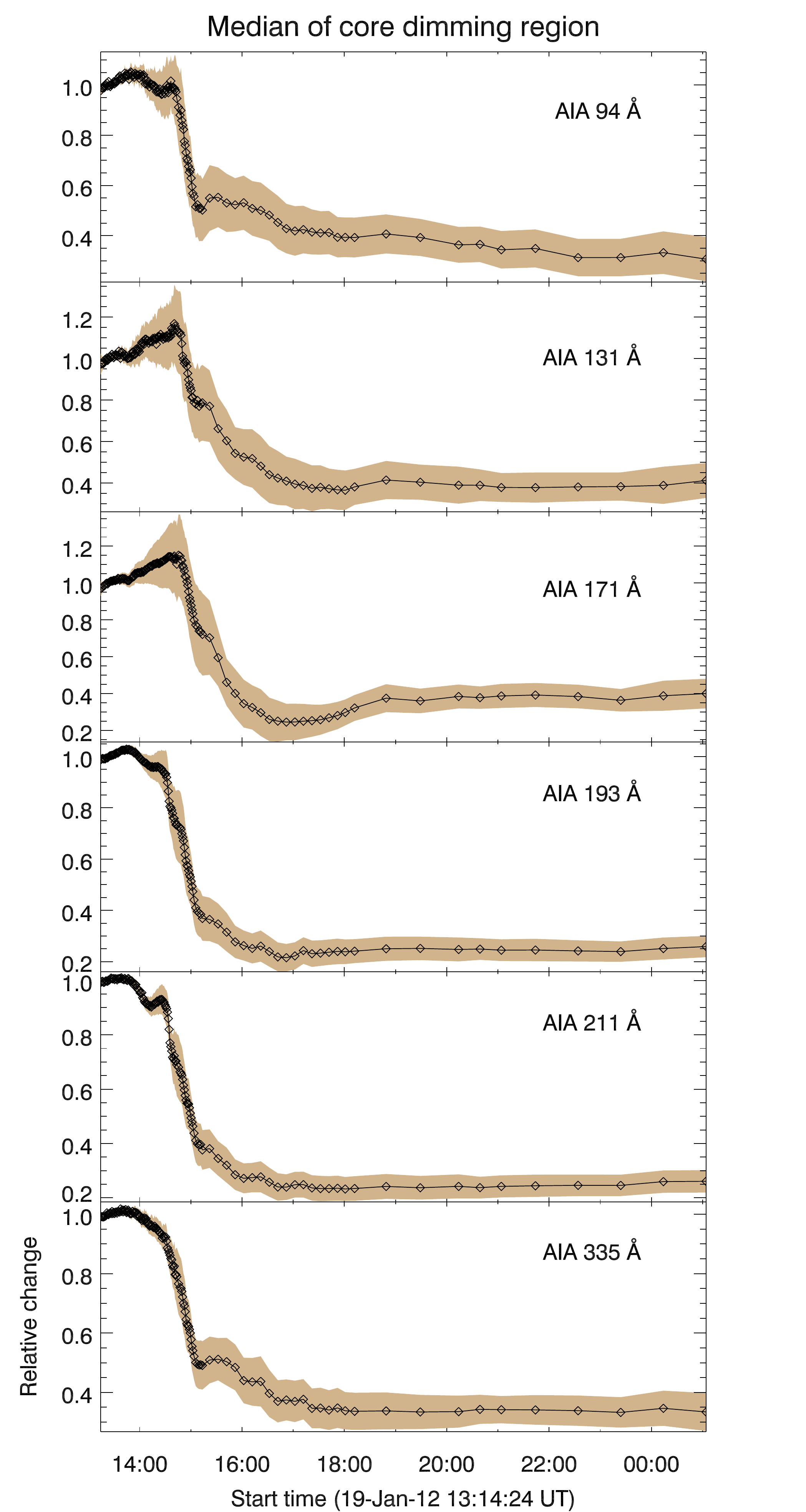}
\end{minipage}\hspace{-1cm}\begin{minipage}[t]{0.4\columnwidth}~
\includegraphics[width=6.1cm]{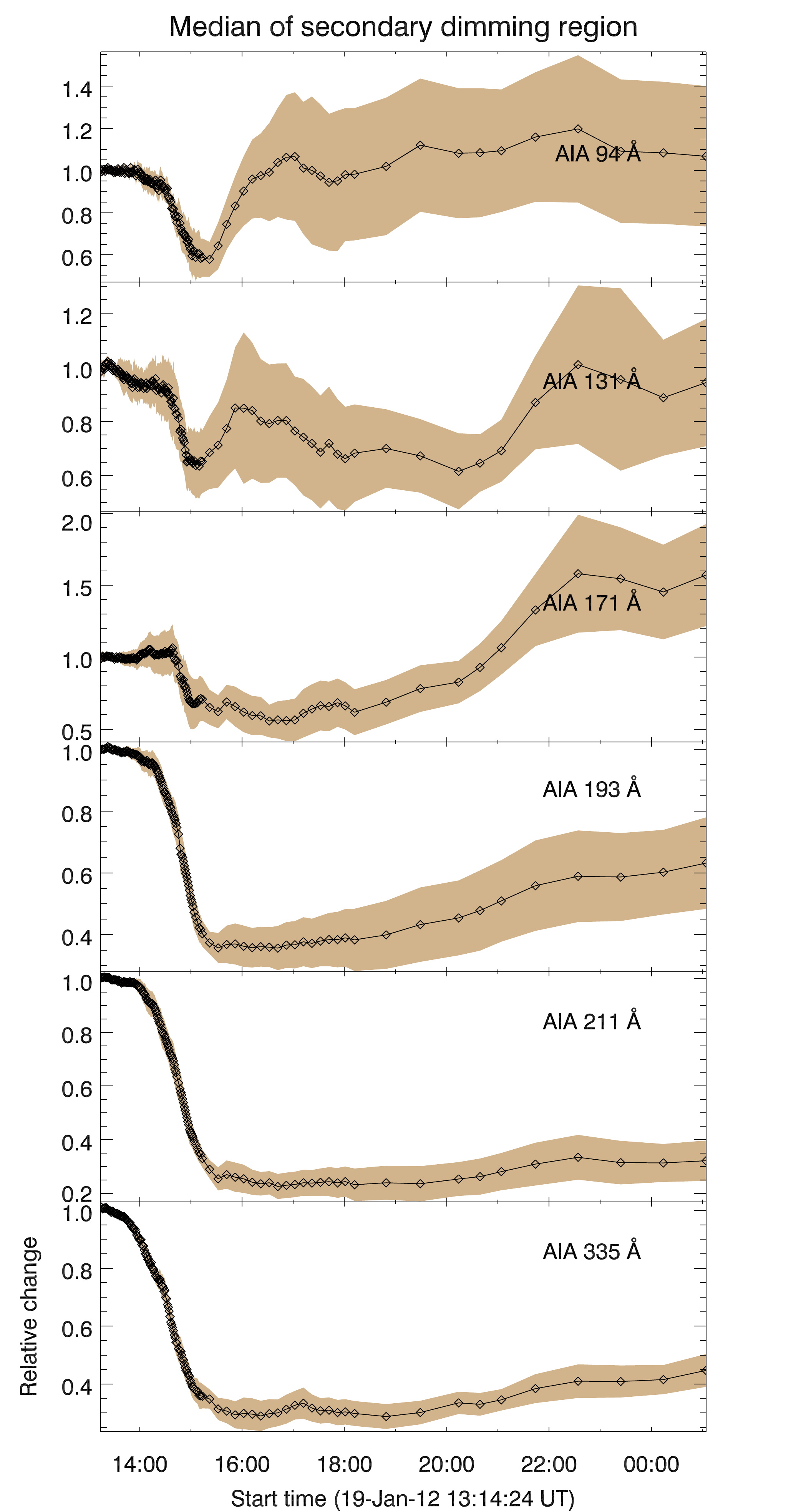}
\end{minipage}
\caption{Same as Figure\,\ref{fig:median_euv_lc} but for the event on 19 January 2012.}
\label{fig:5-median_euv_lc}
\end{figure}

\begin{figure*}
\centering
\vspace{0pt}
\begin{minipage}[b]{0.4\columnwidth}
\includegraphics[width=6.5cm]{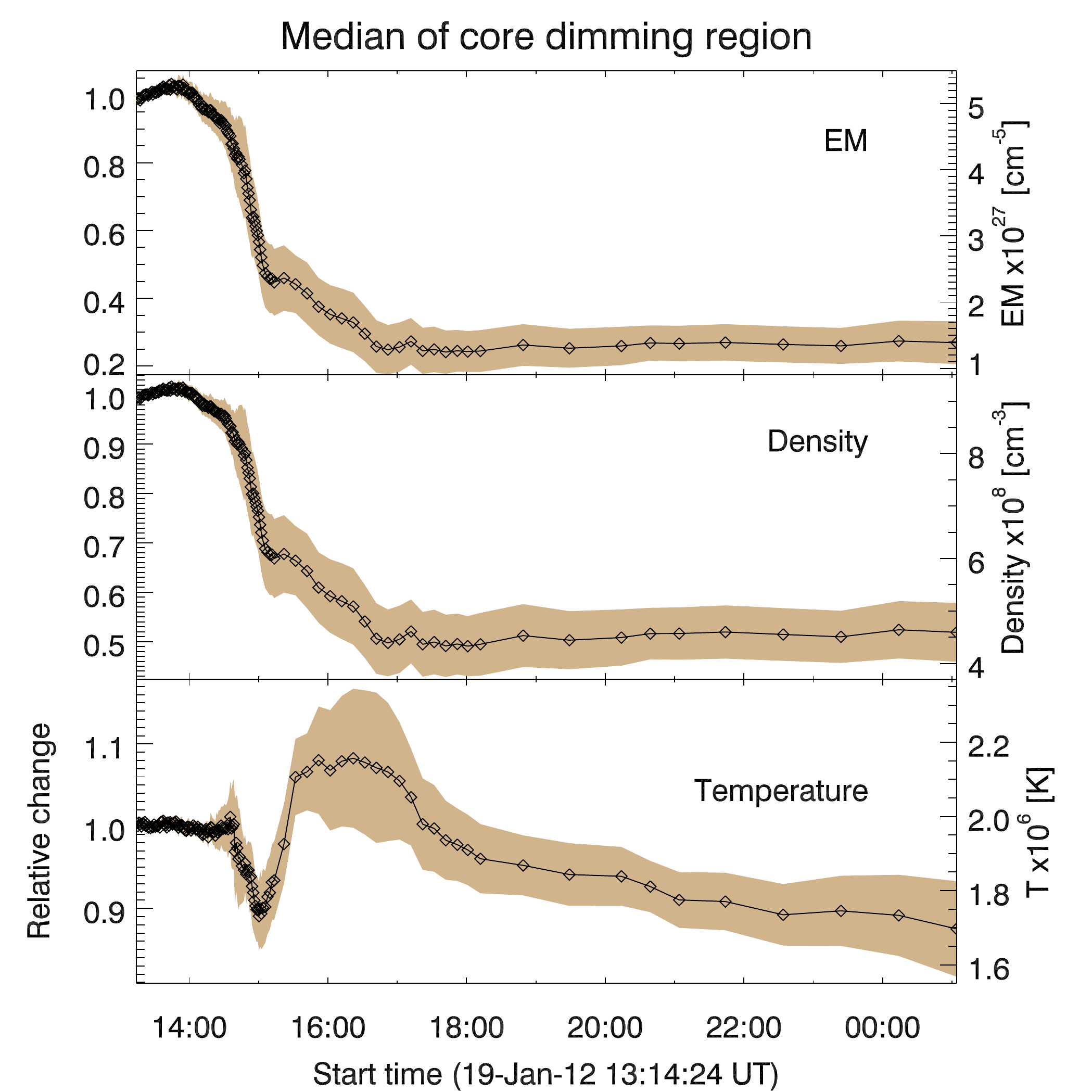}
\end{minipage}\hspace{-1cm}\begin{minipage}[t]{0.4\columnwidth}~
\includegraphics[width=6.5cm]{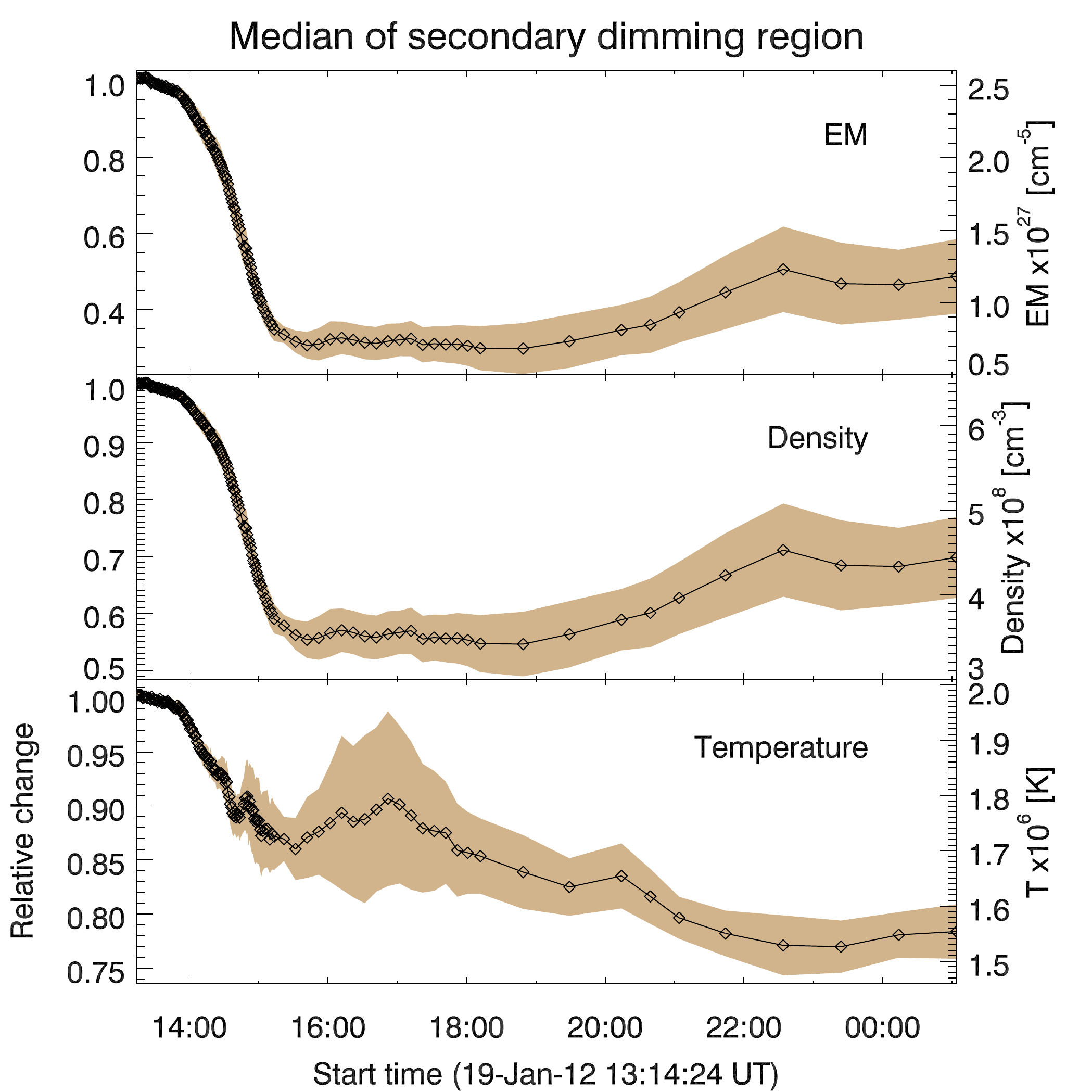}
\end{minipage} 
\caption{Same as Figure\,\ref{fig:median_plasma_lc} but for the event on 19 January 2012.}
\label{fig:5-median_plasma_lc}
\end{figure*}

\clearpage

\subsection{Event on 09 March 2012}
\label{ssec:09Mar2012}

\begin{figure*}[ht]
\centering
\includegraphics[width=\textwidth]{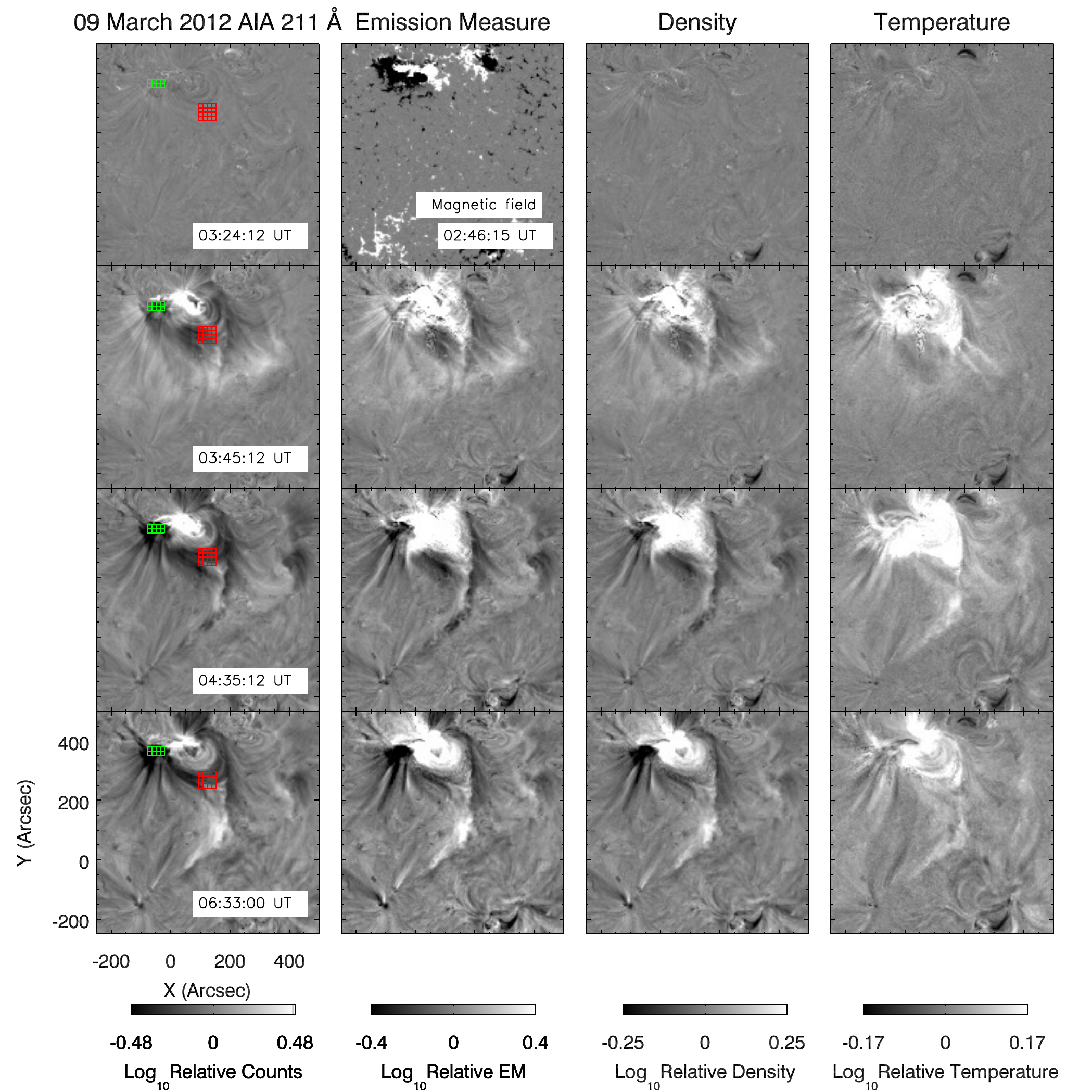}
\vspace{1cm}
\caption{Same as Figure\,\ref{fig:plasma_evol_baseratio} but for the event on 09 March 2012.}
\label{fig:6-plasma_evol_baseratio}
\end{figure*}


\begin{figure}
\centering
\begin{minipage}[t]{0.4\columnwidth}
\includegraphics[width=6.1cm]{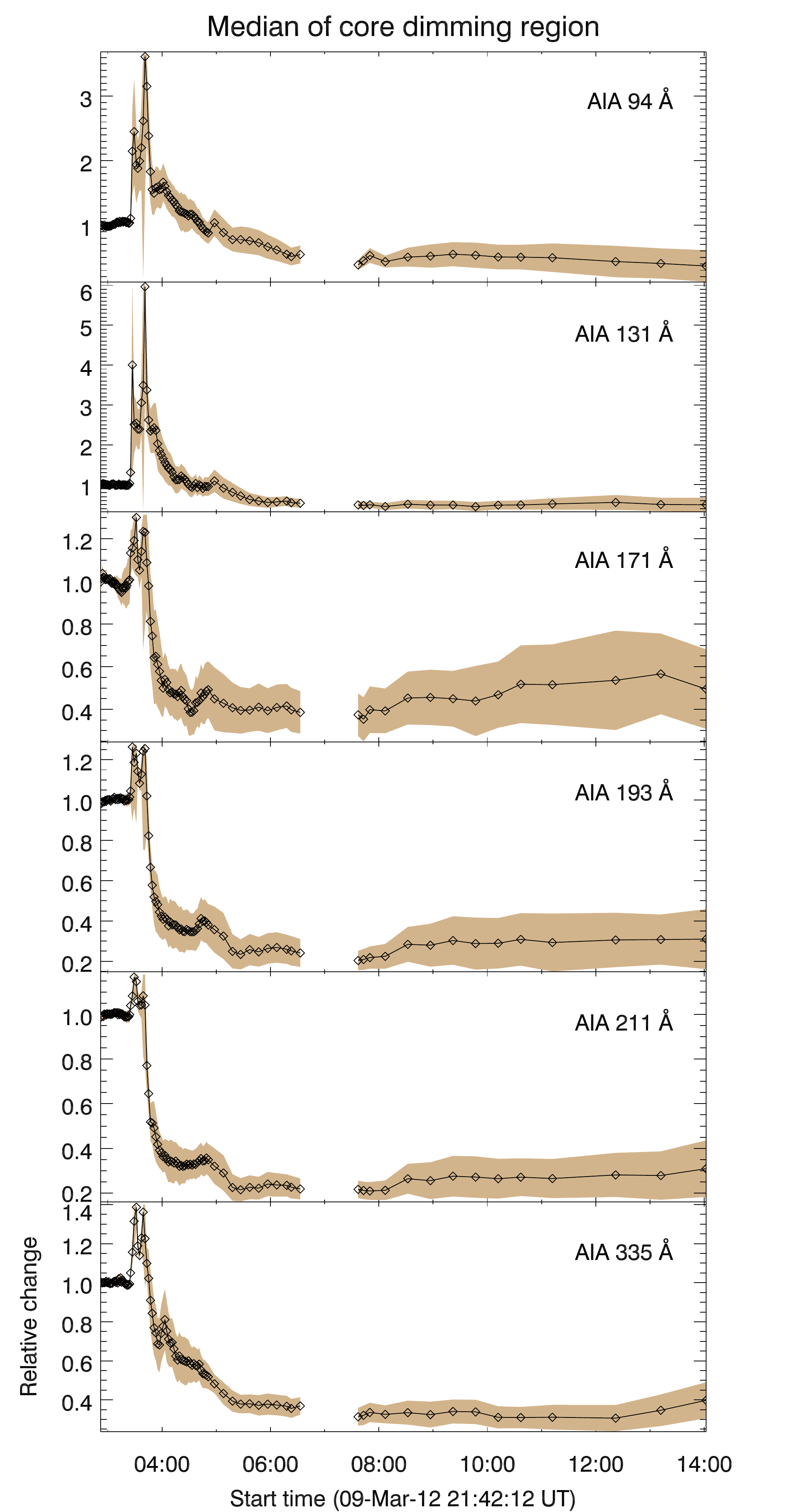}
\end{minipage}\hspace{-1cm}\begin{minipage}[t]{0.4\columnwidth}~
\includegraphics[width=6.1cm]{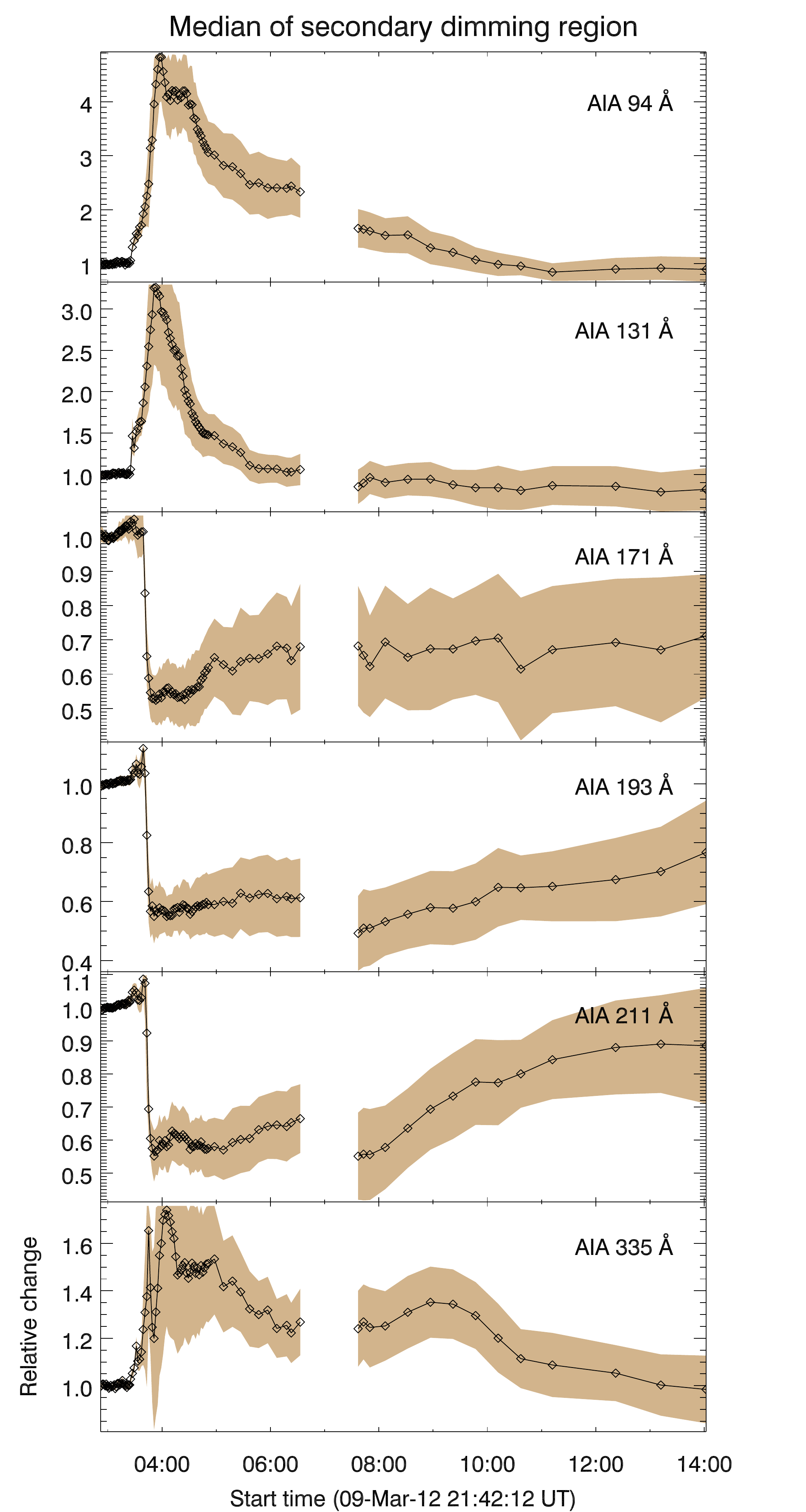}
\end{minipage}
\caption{Same as Figure\,\ref{fig:median_euv_lc} but for the event on 09 March 2012.}
\label{fig:6-median_euv_lc}
\end{figure}

\begin{figure*}
\centering
\begin{minipage}[t]{0.4\columnwidth}
\includegraphics[width=6.5cm]{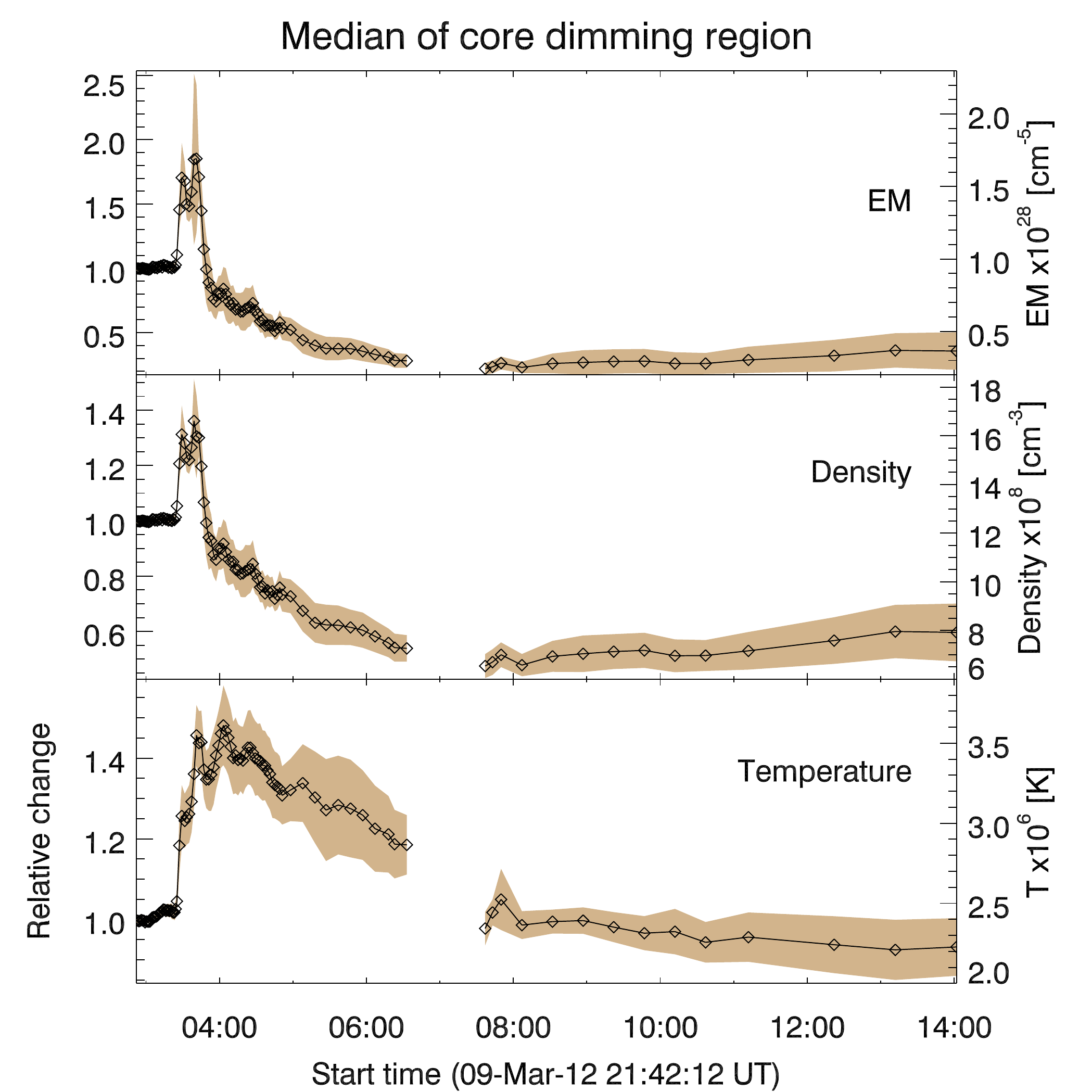}\\
\end{minipage}\hspace{-1cm}\begin{minipage}[t]{0.4\columnwidth}~
\includegraphics[width=6.5cm]{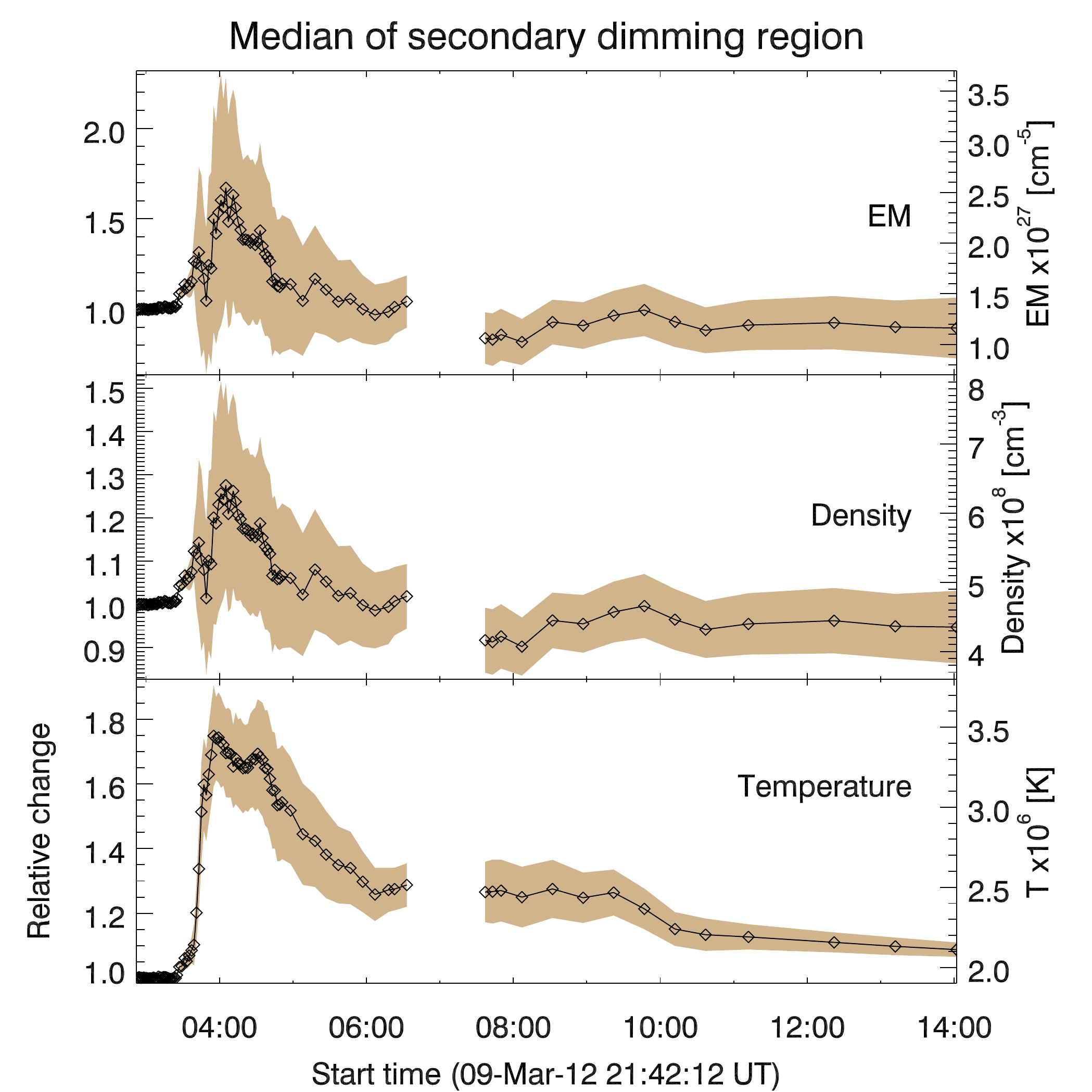}
\end{minipage}
\caption{Same as Figure\,\ref{fig:median_plasma_lc} but for the event on 09 March 2012.}
\label{fig:6-median_plasma_lc}
\end{figure*}

\clearpage



\clearpage
\bibliography{references}



\end{document}